\documentclass[a4paper,reqno]{article}

    \usepackage[english]{babel} 

    \usepackage{graphicx}
    \usepackage[]{color}  
    \usepackage[dvipsnames]{xcolor} 
    \usepackage{csquotes} 
    \usepackage{appendix} 
    \usepackage[a4paper]{geometry} 
    \usepackage{cancel} 
    \usepackage{tikz}
    \usepackage{booktabs}
    \usepackage{enumerate}
    \usepackage{stmaryrd}
        \usetikzlibrary{backgrounds}

    \usepackage{amsmath,amssymb,amsfonts,amsthm} 
    \usepackage{mathtools,todonotes} 
    \usepackage{tikz-cd}  
    \usepackage[all,cmtip]{xy} 
    \usepackage[bbgreekl]{mathbbol} 
    \usepackage{array} 

    \usepackage{hyperref} 
    \usepackage{authblk} 
    \makeatletter
    \newcommand{\subjclass}[2][1991]{%
      \let\@oldtitle\@title%
      \gdef\@title{\@oldtitle\footnotetext{#1 \emph{Mathematics subject classification.} #2}}%
    }
    \newcommand{\keywords}[1]{%
      \let\@@oldtitle\@title%
      \gdef\@title{\@@oldtitle\footnotetext{\emph{Key words and phrases.} #1.}}%
    }
    
\theoremstyle{definition} 
    \newtheorem{definition}{Definition}

\theoremstyle{plain} 
    \newtheorem{theorem}[definition]{Theorem}
    \newtheorem{proposition}[definition]{Proposition}
    \newtheorem{lemma}[definition]{Lemma}
    \newtheorem{corollary}[definition]{Corollary}

\theoremstyle{remark} 
    \newtheorem{remark}[definition]{Remark}

\RequirePackage{tikz-cd}
\RequirePackage{amssymb}
\usetikzlibrary{calc}
\usetikzlibrary{decorations.pathmorphing}

\tikzset{curve/.style={settings={#1},to path={(\tikztostart)
    .. controls ($(\tikztostart)!\pv{pos}!(\tikztotarget)!\pv{height}!270:(\tikztotarget)$)
    and ($(\tikztostart)!1-\pv{pos}!(\tikztotarget)!\pv{height}!270:(\tikztotarget)$)
    .. (\tikztotarget)\tikztonodes}},
    settings/.code={\tikzset{quiver/.cd,#1}
        \def\pv##1{\pgfkeysvalueof{/tikz/quiver/##1}}},
    quiver/.cd,pos/.initial=0.35,height/.initial=0}

\tikzset{tail reversed/.code={\pgfsetarrowsstart{tikzcd to}}}
\tikzset{2tail/.code={\pgfsetarrowsstart{Implies[reversed]}}}
\tikzset{2tail reversed/.code={\pgfsetarrowsstart{Implies}}}
\tikzset{no body/.style={/tikz/dash pattern=on 0 off 1mm}}

\usepackage[bibstyle=alphabetic,citestyle=alphabetic,useprefix,giveninits=true, sorting=ynt, sortcites, minbibnames=99,maxbibnames=99,backend=biber]{biblatex}  
\renewbibmacro{in:}{} 
\bibliography{id on gammas}  
\DeclareSourcemap{
  \maps[datatype=bibtex]{
    \map[overwrite]{ 
      \step[fieldsource=doi, final]
      \step[fieldset=url, null]
      \step[fieldset=eprint, null]
      }
     \map[overwrite]{ 
      \step[fieldsource=eprint, final]
      \step[fieldset=pages, null]
      \step[fieldset=eid, null]
      \step[fieldset=journal, null]
    }  
  }
}

    \DeclarePairedDelimiter\floor{\lfloor}{\rfloor}
    \newcommand{\Tr}[0]{\text{Tr}}
    
    \newcommand{\ugam}{\underline{\gamma}}

    \newcommand{\qsp}[2]{\,\ensuremath{\raise.5ex\hbox{$#1$}\big\slash\raise-.5ex\hbox{$#2$}}}

    \makeatletter
        \newcommand{\zzlabel}[1]{\ifmeasuring@\else\ltx@label{#1}\fi} 
    \makeatother

    \newcounter{terms}[equation] 

\title{Tools for Supergravity in the spin coframe formalism
\thanks{The author acknowledges funding from the EU project Caligola HORIZON-MSCA-2021-SE-01, Project ID: 101086123.}}
\author[]{Filippo Fila-Robattino}

\renewcommand\footnotemark{}

\affil{Institut f\"ur Mathematik, Universit\"at Z\"urich }
\affil{Scuola Internazionale Superiore di Studi Avanzati, Trieste, \href{mailto:ffilarob@sissa.it}{ffilarob@sissa.it}}

\date{}

\keywords{Majorana Spinors. Clifford Algebras. Spin coframe formalism. Supergravity.}
\subjclass[2020]{15A66 (primary); 83C47,  83C470 (secondary).}

\begin{document}
\maketitle

\begin{abstract}
    This paper contains a review of the theoretical foundations of Clifford algebras, spinors and spinor bundles in the so-called co-frame formalism.  A compact index-free notation is introduced, along with a series of identities useful for computations in supergravity theories.
\end{abstract}

\tableofcontents

\section{Introduction}
Spinors are fundamental in the description of supersymmetry theories and, specifically, of supergravities. Historically, mathematicians and physicists have adopted different notations for the same objects, used in different context. In particular, the definition of supergravity theories requires considering different kinds of spinors depending on the spacetime dimensions, which is often a source of confusion to the uninitiated reader.

This review is an attempt to provide a self-contained account on the fundamental tools used in the study of spinors, trying to reconcile the rigorous mathematical definitions with the less precise physics terminology. The secondary scope of these notes is to be a repository of results which are used by the author in forthcoming papers on supergravity, removing the necessity to provide a series of heavy technical proofs, which are conveniently regrouped here. 

The work is organized as follows: the first part of section \ref{clif spin} provides all the necessary definitions and a classification of real and complex Clifford algebras, followed by the definition and main results on spin groups (and their Lie algebras). Furthermore, a systematic classification of the representations of Clifford algebras is presented, with a particular interest in the Lorentzian case, in which we provide a constructive method to obtain the so--called gamma representation, commonly used in physics. The last part of this section is devoted to the definition of Majorana spinors, a central object in the theories of supergravity, showing the direct correlation between the existence of a real structure and the so--called charge conjugation matrix.

In section \ref{sec: spinor fields on manifolds}, we employ the ideas developed in the previous chapter within the context of differential geometry, providing a global description of spin structures and, specifically, spin coframes, a concept which is particularly useful in supergravity. Indeed one can show that the notion of spin coframes is equivalent to that of spin structure (and, in particular, requires the same topological assumptions to exist), with the advantage of providing a framework which allows to define spinor fields without the necessity of fixing a metric, which is ultimately considered as a dynamical object in the context of physics.

Lastly, section \ref{sec: identities} contains some very well known identities, as well as some lesser known ones. A full description on how to obtain Fierz rearrangements in $D=4$ is presented, with a particularly useful example in the mostly plus Lorentzian signature. Most of these results are rephrased in the index--free notation provided by the spin coframe formalism. Finally, the last part presents a series of technical lemmata in dimension 4, which, as previously anticipated, acts as a repository of results useful in future works of the author. 


\section{Clifford algebras and spin groups}\label{clif spin}

Some of the introductory content in the following section has appeared in \cite{Canepa:2022uaq} and has been reported to have a complete discussion with a consistent notation. The remaining section on Clifford algebras mainly follows ,\cite{Kostant1987},\cite{fatibene2018}, \cite{spingeom}, \cite{RudolphSchmid} and \cite{Figueroa}.

The parts regarding Majorana spinors and Fierz identities follow \cite{Figueroa}, \cite{Castellani:1991et}, \cite{Freedman:2012zz}, \cite{KUGO1983357}, \cite{Scherk:1978fh} and \cite{Dabrowski:88}. References \cite{harvey1990spinors, VanNieuwenhuizen:1985be} are also recommended.

\subsection{Clifford algebras}
Let $V$ be a real vector space of dimension $D$ with an inner product of signature $(r,s)$. Let $\eta_{ab}$ be the matrix diag$(-1,\cdots,-1,1,\cdots,1)$ with $r$ plus 1 and $s$ minus 1, giving the inner product on $V$ with respect to an orthonormal basis $\{v_a\}$, $a=1,\cdots,D$. 

We define the Clifford algebra on $V$ by means of its universal property. In particular	
\begin{definition}[Clifford map]
	A \text{Clifford map} is given by the pair $(A,\phi)$ where $A$ is an associative algebra with unity and $\phi$ is a linear map $\phi \colon V\rightarrow A$ such that $\forall u,v\in V$
	\begin{equation}\label{id: Clif condition}
		\phi(u)\phi(u)=-\eta(u,u)\mathbb{1}_A
	\end{equation}
\end{definition}

\begin{definition}[Clifford algebra]
	The Clifford algebra $\mathcal{C}(V)$ is an associative algebra with unit together with a Clifford map $i\colon V\rightarrow \mathcal{C}(V)$ such that any Clifford map factors through a unique algebra homomorphism from $C(V)$. In other words, given any Clifford map $(A,\phi)$ there is a unique algebra homomorphism $\Phi\colon \mathcal{C}( V) \rightarrow A$ such that $\phi = \Phi \circ i$
	\begin{equation*}
		\begin{tikzcd}
			V \arrow[r, "\phi"] \arrow[d, "i"'] & A \\
			\mathcal{C}(V) \arrow[ru, "\Phi"']            &  
		\end{tikzcd}
	\end{equation*}
\end{definition}

\begin{proposition}
    The Clifford algebra of $V$ is unique up to isomorphisms.
\end{proposition}

We give a model for such an algebra. Consider the tensor algebra $T(V):=\mathbb{R}\oplus V \oplus V^{\otimes 2} \oplus \cdots$ and quotient it out  by the two-sided ideal $I(V)$ generated by $v\otimes v + \eta(v,v)\mathbb{1}$, i.e.
\begin{equation*}
    \mathcal{C}(V):=\frac{T(V)}{I(V)}.
\end{equation*}

Indeed one can set $i$ to be the composition of the canonical projection $\rho:T(V)\rightarrow \mathcal{C}(V)$ with the inclusion $V\hookrightarrow T(V)$. Every linear map $\phi: V \rightarrow A$ extends uniquely to an algebra homomorphism $\tilde{\Phi}:T(V)\rightarrow A$, which identically vanishes on $I(V)$ by \eqref{id: Clif condition}. This implies that $\tilde{\Phi}$ uniquely descends to a homomorphism $\Phi:\mathcal{C}(V)\rightarrow A$, satisfying
    \begin{equation*}
        \Phi\circ i = \phi.
    \end{equation*}

Notice that $T(V)$ is a $\mathbb{Z}$-graded algebra. The ideal $I(V)$ is spanned by elements that are not necessarily homogeneous, therefore the $\mathbb{Z}$--grading is lost in the Clifford algebra. However, the generators of $I(V)$ are even, therefore $\mathcal{C}(V)$ will be $\mathbb{Z}_2$-graded. In particular, it splits into
\begin{equation*}
	\mathcal{C}(V)=\mathcal{C}_0(V) \oplus \mathcal{C}_1(V) .
\end{equation*}
Another important property, for any two vectors $v,w \in V$, is the following
\begin{equation*}
	\begin{split}
		(v+w)^2& =v^2+ vw + wv + w^2 = - \eta(v,v)\mathbb{1} - \eta(w,w)\mathbb{1} + \{v,w\}\\
		& = - \eta(v+w,v+w)\mathbb{1}=-\eta(v,v)\mathbb{1}  - \eta(w,w)\mathbb{1} -2\eta(v,w)\mathbb{1}\\
		\Rightarrow&\quad \{v,w\}:=vw+wv=-2\eta(v,w)\mathbb{1}.
	\end{split}
\end{equation*}
Now, considering an orthonormal basis $\{v_a\}$ of $V$, setting the first $s$ elements $\{v_A\}$ such that $\eta(v_A,v_A)=-1$ and the second $r$ elements $\{v_i\}$ such that $\eta(v_i,v_i)=1$, we obtain $\{v_a,v_b\}=-2\eta_{ab}\mathbb{1}$. This means that when $a\neq b$, $v_av_b=-v_bv_a$ and that $v_av_a=\pm \mathbb{1}$.

At this point, since every element in the tensor algebra $T(V)$ is a finite linear combination of the product of finite elements in the basis of $V$, to obtain elements in $\mathcal{C}(V)$ we simply apply the constraint $\{v_a,v_b\}=-2\eta_{ab}\mathbb{1}$. Indeed, since the elements of $V$ are multiplicative generators of $T(V)$, they must also generate $\mathcal{C}(V)$, hence a basis of Clifford algebra is given in the form
\begin{equation}
\label{Clifford basis}
	\mathbb{1} \quad v_a \quad v_{ab}:=\underset{a<b}{v_a v_b} \quad v_{abc}:=\underset{a<b<c}{v_a v_b v_c} \quad \cdots \quad v_*:=v_1\cdots v_{D}
\end{equation}
The $\mathbb{Z}_2$-grading is now clearer, as we can interpret even (odd) elements of $\mathcal{C}(V)$ to be finite linear combinations of products of an even (odd) number of elements of the basis $V$. In particular, the even part $\mathcal{C}_0(V)$ is a sub-algebra of $\mathcal{C}(V)$, while the odd part $\mathcal{C}_1(V)$ is not (it does not contain the unity). They are both $2^{d-1}$-dimensional, making $\mathcal{C}(V)$ $2^d$-dimensional.

\begin{proposition}
    There exists a canonical isomorphism between the Clifford algebra and the exterior algebra of $V$
    \begin{equation}
        \sigma \colon\mathcal{C}(V) \rightarrow \wedge^\bullet V
    \end{equation}
\end{proposition}
\begin{proof}
    For any $u=u^a v_a \in V$, consider the dual vector $\underline{u} = \eta_{ab}u^b \nu^a$, where $\{\nu^a\}$ is a basis of covectors such that $\nu^a(v_b)=\delta^a_b$. Let $\theta$ be the mapping 
        \begin{equation*}
            \theta\colon V \rightarrow \mathrm{End}(\wedge^\bullet V) \qquad \mathrm{s.t.} \qquad\theta(u)(\alpha)=u\wedge \alpha + \iota_{\underline{u}}\alpha,
        \end{equation*}\qquad 
    where $\iota_{\underline{u}}\alpha$ is the contraction with the covector of $u$ for all $\alpha\in\wedge^\bullet V$. Then one finds
        \begin{align*}
            \theta(u)^2\alpha&= u \wedge u \wedge \alpha + u \wedge \iota_{\underline{u}}\alpha + \iota_{\underline{u}}(u\wedge \alpha) + \iota_{\underline{u}}\iota_{\underline{u}}\alpha\\
            &=u \wedge \iota_{\underline{u}}\alpha + \iota_{\underline{u}}u \wedge \alpha - u\wedge \iota_{\underline{u}}\alpha = \iota_{\underline{u}}u  \alpha \\
            &= \eta(u,u) \alpha.
        \end{align*}
    This implies, by the universal property, the existence of an algebra morphism
        \begin{equation*}
            \hat{\theta}:\mathcal{C}(V) \rightarrow \mathrm{End}(\wedge^\bullet V),
        \end{equation*}
    which, composed with with the identity element in $\mathrm{End}(\wedge^\bullet V)$, yields
        \begin{equation*}
            \sigma:\mathcal{C}(V) \rightarrow \wedge^\bullet V.
        \end{equation*}
    It is immediate to check that an element $u_1\cdots u_k\in\mathcal{C}(V)$ is sent to $u_1\wedge\cdots\wedge u_k\in\wedge^\bullet V$, hence one obtains that a basis of $\mathcal{C}(V)$ is sent to a basis of $\wedge^\bullet V$, proving that $\sigma$ defines an isomorphism.   

    \begin{remark}
        The highest grade basis element $v_*$ is also known as volume element, in analogy with its image under $\sigma$, defining the volume form on $V$.
    \end{remark}
\end{proof}

\subsection{Classification of Clifford Algebras}
We start by classifying real Clifford algebras. In this section, we denote by $\mathcal{C}(r,s)$ the Clifford algebra over the $D$-dimensional real vector space $V$ endowed with a non-degenerate metric of signature $(r,s)$. We will also denote by $\mathbb{K}(N)$ the $N\times N$ matrices over the field $\mathbb{K}$, while, in view of the future definition of gamma matrices, in this section we will denote the generators of the Clifford algebra by $\Gamma_a$. We first consider the low-dimensional Clifford algebras, which will provide the fundamental building blocks to obtain the higher dimensional ones.

\begin{lemma}\label{lem: low dim Clif alg}
    \begin{align*}
        (\mathrm{i})\hspace{1mm} \mathcal{C}(1,0)&\simeq \mathbb{C}, \quad(\mathrm{ii})\hspace{1mm} \mathcal{C}(0,1)\simeq \mathbb{R}\oplus \mathbb{R}, \quad (\mathrm{iii})\hspace{1mm}\mathcal{C}(1,1)\simeq \mathbb{R}(2),\\
        (\mathrm{iv})\hspace{1mm}&\mathcal{C}(0,2)\simeq \mathbb{R}(2), \quad (\mathrm{v})\hspace{1mm}\mathcal{C}(2,0)\simeq \mathbb{H}.
    \end{align*}
\end{lemma}
\begin{proof}
    In order to prove the above statements, we pick a representation of the Clifford algebra in terms of matrices, 
    \begin{enumerate}[(i)]
        \item there is only one element $\{v_1\}$ in the basis of $V$, such that $\Gamma^2_1=-\mathbb{1}$, defining  a complex structure on $T(V)$, hence $\mathcal{C}(1,0)=\mathbb{C}$;
        \item analogously, we find $\Gamma_1^2=\mathbb{1}$, hence $\mathcal{C}(0,1)=\mathbb{R}\oplus \mathbb{R}$;
        \item following the physics notation and setting $\{v_0,v_1\}$ as basis of $V$ such that $\Gamma_0^2=\mathbb{1}$ and $\Gamma_1^2=-\mathbb{1}$, we can choose the following anticommuting matrices
            \begin{equation}
                \Gamma_0=\sigma_1=\begin{pmatrix}
                    0 & 1 \\
                    1 & 0
                \end{pmatrix}\qquad \text{and} \qquad \Gamma_1=i\sigma_2 = \begin{pmatrix}
                    0 & 1\\
                    -1 & 0
                \end{pmatrix},
            \end{equation}
        They are $2\times 2$ real matrices, hence they generate $\mathcal{C}(1,1)=\mathbb{R}(2)$. The even part is generated by $\mathbb{1}$ and $\Gamma_*=\Gamma_0 \Gamma_1$, given by 
            \begin{equation*}
                \Gamma_*=-\sigma_3 =\begin{pmatrix}
                    -1 & 0 \\
                    0 & 1
                \end{pmatrix},
            \end{equation*}
        hence obtaining $\mathcal{C}_0(1,1)$ as the diagonal $2\times 2$ real matrices;
        \item in the case of $\mathcal{C}(0,2)$, we pick anticommuting matrices $\Gamma_1$ and $\Gamma_2$ squaring to $\mathbb{1}$, which are explicitly realized by
            \begin{equation*}
                \Gamma_1=\sigma_1=\begin{pmatrix}
                    0 & 1 \\
                    1 & 0
                \end{pmatrix}\qquad \text{and} \qquad \Gamma_2=\sigma_3 = \begin{pmatrix}
                    1 & 0\\
                    0 & -1
                \end{pmatrix},
            \end{equation*}
        as before we obtain $\mathcal{C}(0,2)=\mathbb{R}(2)$, and the volume element is given by    
            \begin{equation*}
                \Gamma_*=-i\sigma_2=\begin{pmatrix}
                    0 & -1\\
                    1 & 0
                \end{pmatrix},
            \end{equation*}
        which defines a complex structure as it squares to $-\mathbb{1}$. Hence the even subalgebra, being generated by $\mathbb{1}$ and $\Gamma_*$, is $\mathcal{C}_0(0,2)=\mathbb{C}$;
        \item for $\mathcal{C}(2,0)$ we need two anticommuting matrices  $\Gamma_1$ and $\Gamma_2$ squaring to $-\mathbb{1}$, which are explicitly realized by
            \begin{equation*}
                \Gamma_1=i\sigma_1=\begin{pmatrix}
                    0 & i \\
                    i & 0
                \end{pmatrix}\qquad \text{and} \qquad \Gamma_2=i\sigma_2 = \begin{pmatrix}
                    0 & 1\\
                    -1 & 0
                \end{pmatrix},
            \end{equation*}
        both squaring to $\mathbb{-1}$, hence defining two anticommuting complex structures. The Clifford algebra then has to coincide with the algebra of quaternions $\mathbb{H}$, explicitly realized by the identification
            $\{1,i,j,k\}=\{\mathbb{1},\Gamma_1,\Gamma_2,\Gamma_*\}$, where
            \begin{equation*}
                \Gamma_*=-i\sigma_3=\begin{pmatrix}
                    -i & 0\\
                    0 & i
                \end{pmatrix},
            \end{equation*}
        The even subalgebra is generated by $\mathbb{1}$ and $\Gamma_*$, which again defines a complex structure, hence obtaining $\mathcal{C}_0(0,2)=\mathbb{C}$.    
    \end{enumerate}
\end{proof}

The following lemma allows to recover higher dimensional Clifford algebras from the lower dimensional ones

\newpage
\begin{lemma}\label{lem: clif alg low to high dim}
    The following statements are true
        \begin{enumerate}[(i)]
            \item $\mathcal{C}(d,0)\otimes \mathcal{C}(0,2)\simeq \mathcal{C}(0,d+2)$
            \item $\mathcal{C}(0,d)\otimes \mathcal{C}(2,0)\simeq \mathcal{C}(d+2,0)$
            \item $\mathcal{C}(r,s)\otimes\mathcal{C}(1,1)\simeq\mathcal{C}(r+1,s+1)$
        \end{enumerate}
\end{lemma}
\begin{proof}
    For $(i)$, consider $\{v_i\}$, $i=1,\cdots,d$ and $\{v_\alpha\}$, $\alpha=d+1,d+2$ respectively generating  $\mathcal{C}(d,0)$ and $\mathcal{C}(0,2)$. Then there are relations
        \begin{equation*}
            v_i \cdot v_j = -2 \delta_{ij}\mathbb{1} \qquad \text{and}\qquad v_\alpha\cdot v_\beta = 2\delta_{\alpha\beta}.
        \end{equation*}
    We can define new elements $\{v_a\}$
, $a=1,\cdots ,d+2$ as
        \begin{equation*}
            v_a:=\begin{cases}
                &v_i \otimes v_{d+1}\cdot v_{d+2} \quad a\leq d\\
                &\mathbb{1}\otimes v_\alpha \qquad  \qquad \quad a > d
            \end{cases}
        \end{equation*}
    A quick computation gives $$v_a\cdot v_b=2\delta_{ab}\mathbb{1},$$ hence proving the $v_a$'s generate $\mathcal{C}(0,d+2)$. 

    The case of $(ii)$ is analogous. For $(iii)$ consider $\{v_1,\cdots,v_r,v_{r+1},\cdots,v_{r+s}\}$ as a basis of $\mathbb{R}^{r,s}$, generating $\mathcal{C}(r,s)$, and $\{v'_1,v'_2\}$ as generating $\mathcal{C}(1,1)$. Then we define a new set of vectors $\{v_a\}$, $a=1,\cdots,d+2$ such that
        \begin{equation*}
            v_a=\begin{cases}
                &v_a \otimes v'_1\cdot v'_2, \quad 1\leq a\leq r \\
                & \mathbb{1}\otimes v'_1 \qquad  \qquad \quad a = r+1\\
                & v_{a-1} \otimes  v'_1\cdot v'_2 \qquad r+1\leq a \leq d+1\\
                &\mathbb{1}\otimes v'_2 \qquad  \qquad \quad a = d+2
            \end{cases}
        \end{equation*}
    A quick computation shows that the newly defined $v_a's$ generate $\mathcal{C}(r+1,s+1)$.
\end{proof}

As a result, one can show that structure of the $(r,s)$ real Clifford algebra has periodicity 8 in $r-s$. The following prposition allows us to classify the even Clifford subalgebras.

\begin{proposition}\label{prop: even Clif}
    The even Clifford subalgebra is related to the full one in the following way
    \begin{equation}
        \mathcal{C}_0(r+1,s)\simeq \mathcal{C}(s,r) \qquad \text{and}\qquad \mathcal{C}_0(r,s+1)\simeq \mathcal{C}(r,s),
    \end{equation}
    furthermore,
    \begin{equation}
        \mathcal{C}_0(r,s)\simeq \mathcal{C}_0(s,r).
    \end{equation}
\end{proposition}

\newpage
Taking into account the periodicity of the structure of Clifford algebras, we obtain the following classification 
\begin{table}[h]\label{table: Clif alg}
\centering
\begin{tabular}{@{}lll@{}}
\toprule
$r-s$ mod 8 & $\mathcal{C}(r,s)$                  & $N$                     \\ \midrule
0,6         & $\mathbb{R}(2^\frac{N}{2})$                     & ${D}$     \\
2,4         & $\mathbb{H}(2^\frac{N}{2})$                     & $D-2$ \\
1,5         & $\mathbb{C}(2^\frac{N}{2})$                     & $D-1$ \\
3           & $\mathbb{H}(2^\frac{N}{2})\oplus \mathbb{H}(2^\frac{N}{2})$ & $D-3$ \\
7           & $\mathbb{R}(2^\frac{N}{2})\oplus\mathbb{R}(2^\frac{N}{2})$  & $D-1$ \\ \bottomrule
\end{tabular}
\qquad 
\begin{tabular}{@{}lll@{}}
\toprule
$r-s$ mod 8 & $\mathcal{C}_0(r,s)$                  & $N$                     \\ \midrule
1,7         & $\mathbb{R}(2^\frac{N}{2})$                     & ${D-1}$     \\
3,5         & $\mathbb{H}(2^\frac{N}{2})$                     & $D-3$ \\
2,6         & $\mathbb{C}(2^\frac{N}{2})$                     & $D-2$ \\
4           & $\mathbb{H}(2^\frac{N}{2})\oplus \mathbb{H}(2^\frac{N}{2})$ & $D-4$ \\
0           & $\mathbb{R}(2^\frac{N}{2})\oplus\mathbb{R}(2^\frac{N}{2})$  & $D-2$ \\ \bottomrule
\end{tabular}
\caption{Clifford algebras and even Clifford subalgebras in various dimensions} 
\end{table}

The situation is significantly simplified when one takes into consideration the complexification of the Clifford algebras. Consider $V_\mathbb{C}=V\otimes_{\mathbb{R}} \mathbb{C}$ and define the mapping
    \begin{equation*}
        \hat{i}\colon V\otimes_{\mathbb{R}} \mathbb{C} \rightarrow \mathcal{C}(V)\otimes_{\mathbb{R}} \mathbb{C} \colon u\otimes z \mapsto i(u)\otimes z,
    \end{equation*}
then $\hat{i}(u\otimes z)^2=i(u)^2\otimes z^2 = -\eta(u,u)\mathbb{1}\otimes z^2 = -\eta(u\otimes z,u\otimes z) \mathbb{1}$, proving that 
    \begin{equation}\label{eq:cplx clif}
        \mathcal{C}(V)_\mathbb{C}=\mathcal{C}(V)\otimes_{\mathbb{R}} \mathbb{C}=\mathcal{C}(V_\mathbb{C}).
    \end{equation}

Now, since on $V_\mathbb{C}$ it is always possible to diagonalize $\eta$ to a Euclidean metric, denoting by $\mathcal{C}(D)$ the complex Clifford algebra over $\mathbb{C}^D$, one obtains
    \begin{equation}\label{eq: cplx Clif to real}
        \mathcal{C}(D)\simeq \mathcal{C}(D,0)_\mathbb{C}\simeq \mathcal{C}(D-1,1)_\mathbb{C}\simeq\cdots\simeq \mathcal{C}(0,D)_\mathbb{C}.
    \end{equation}

Notice also that the above statement, together with proposition \ref{prop: even Clif}, implies that 
    \begin{equation}\label{eq: cplx even clif to full}
        \mathcal{C}_0(D)\simeq \mathcal{C}(D-1).
    \end{equation}

\begin{proposition}\label{prop: complex clifford}
    \begin{equation}
        \mathcal{C}(n+2)\simeq \mathcal{C}(n)\otimes \mathbb{C}(2), \qquad \mathcal{C}(2k)\simeq \mathbb{C}(2^k), \qquad \mathcal{C}(2k+1)\simeq \mathbb{C}(2^k)\oplus \mathbb{C}(2^k).
    \end{equation}
\end{proposition} 
\begin{proof}
    Using lemma \ref{lem: clif alg low to high dim} and eq. \eqref{eq: cplx Clif to real}, we see that
        \begin{equation*}
            \mathcal{C}(n+2)\simeq (\mathcal{C}(n,0)\otimes_\mathbb{R}\mathbb{C})\otimes_\mathbb{C} (\mathcal{C}(0.2)\otimes_\mathbb{R}\mathbb{C})\simeq\mathcal{C}(n)\otimes_\mathbb{C}\mathcal{C}(2).
        \end{equation*}
    By lemma \ref{lem: low dim Clif alg} and \eqref{eq:cplx clif} we obtain 
        \begin{align*}
            \mathcal{C}(1)\simeq \mathbb{C}\oplus\mathbb{C} \qquad \mathrm{and} \qquad \mathcal{C}(2)\simeq \mathbb{C}(2), 
        \end{align*}
    thanks to which, by iteration of the above result, we obtain
        \begin{equation*}
            \mathcal{C}(2k)\simeq\bigotimes^k \mathbb{C}(2)\simeq \mathrm{End}(\bigotimes^k \mathbb{C}^2)\simeq \mathbb{C}(2^k)
        \end{equation*}
    and    
        \begin{equation*}
            \mathcal{C}(2k+1)\simeq\bigotimes^k \mathbb{C}(2)\oplus \bigotimes^k \mathbb{C}(2) \simeq\mathbb{C}(2^k)\oplus \mathbb{C}(2^k).
        \end{equation*}
\end{proof}

Therefore, proposition \eqref{eq: cplx even clif to full}, allows to obtain the following classification    
\begin{table}[h]
\centering
\begin{tabular}{@{}lll@{}}
\toprule
$D$ mod 2 & $\mathcal{C}(D)$                  & $N$                     \\ \midrule
0         & $\mathbb{C}(2^\frac{N}{2})$                     & ${D}$     \\
1         & $\mathbb{C}(2^\frac{N}{2})\oplus\mathbb{C}(2^\frac{N}{2})$                     & $D-1$ \\
 \bottomrule
\end{tabular}\qquad
\begin{tabular}{@{}lll@{}}
\toprule
$D$ mod 2 & $\mathcal{C}_0(D)$                  & $N$                     \\ \midrule
0         & $\mathbb{C}(2^\frac{N}{2})\oplus\mathbb{C}(2^\frac{N}{2})$                     & ${D-2}$     \\
1         & $\mathbb{C}(2^\frac{N}{2})$                     & $D-1$ \\
 \bottomrule
\end{tabular}
\label{table: complex Clif clas}
\caption{Complex Clifford algebra and even subalgebra in various dimensions} 
\end{table}

\subsection{Pin and Spin groups}

\begin{definition}[grading map]
	Consider the Clifford map $i\colon V\rightarrow \mathcal{C}(V)$. By abuse of notation, this map sends $v$ to $v$ inside $\mathcal{C}(V)$. Defining $\alpha:=-i\colon v\rightarrow \mathcal{C}(V):v\mapsto -v$, it has the property that $\alpha(v)\alpha(v)=-\eta(v,v)\mathbb{1}$. We can extend it to the whole $\mathcal{C}(V)$ as $\alpha\colon \mathcal{C}(V)\rightarrow \mathcal{C}(V)$ by restricting it to the identity on even elements, to minus the identity on odd elements. This map is called \text{grading} (or parity) since it essentially defines the $\mathbb{Z}_2$-grading on $\mathcal{C}(V)$.
\end{definition}
Clearly we have that $\alpha\circ\alpha=\mathbb{1}$, therefore $\alpha$ is invertible and equal to its inverse.

\begin{definition}[transpose]
	Let $S=u_1 u_2\cdots u_k\in \mathcal{C}(V)$. We define the \text{transpose} of $S$ to be
	\begin{equation*}
		{}^t(S)={}^t(u_1 u_2\cdots u_k):=u_k\cdots u_2 u_1=:\overline{S}
	\end{equation*}
	It is well defined since the generators of the Clifford ideal are invariant under the transposition.
\end{definition}
Furthermore, the transpose preserves the grading, namely ${}^t(\alpha(S))=\alpha(^t(S))$.

It is a well known fact that not all elements in $\mathcal{C}(V)$ are invertible. Let us define the multiplicative subgroup $\mathcal{C}^*(V)\subset \mathcal{C}(V)$ of invertible elements. Clearly every subgroup of $\mathcal{C}(V)$ is contained in $\mathcal{C}^*(V)$. 

\begin{definition}[Clifford group]
    The Clifford group is defined to be the Lie subgroup of $\mathcal{C}^*(V)$, given by
        \begin{equation*}
            \Gamma(V):=\{ S\in\mathcal{C}^*(V) \hspace{1mm}\vert\hspace{1mm}\forall u\in V, \alpha(S)uS^{-1}\in V \}.
        \end{equation*}
    The map $l:\Gamma(V)\rightarrow \mathrm{Aut}(V)$ defined by $\alpha(S)(u)=\alpha(S)uS^{-1}$  is by definition a representation of $\Gamma(V)$, called twisted adjoint representation.
\end{definition}

\begin{lemma}
The twisted adjoint representation is such that

    \begin{enumerate}
        \item $l(\alpha(S))=l(S)$ for all $S\in\Gamma(V)$;
        \item \label{lem: reflections}for any vector $v\in V$ such that $\eta(v,v)=\pm 1$,  the map $l(v)$ is a reflection about the plane orthogonal to the unit vector $v$;
        \item \label{lem: ker l}$\ker(l)\simeq \mathbb{R}^*$.
    \end{enumerate}
\end{lemma}
\begin{proof} We prove each point separately:
    \begin{enumerate}
        \item $l(S)(u)=-\alpha(l(S)(u)=-\alpha(\alpha(S)uS^{-1})=S u \alpha(S)^{-1}=l(\alpha(S)(u))$.
        \item Recalling that $vv=-\eta(v,v)\mathbb{1}=-|v|^2\mathbb{1}$, we have $v^{-1}=-\frac{v}{|v|^2}$. For all $w\in V$, denote $w^\parallel:=\frac{\eta(v,w)}{\eta(v,v)}v$ to be the component of $w$ parallel to $v\in V$. The perpendicular component is defined as $w^\perp := w - w^\parallel$. Then
        	\begin{equation*}
        		\begin{split}
        			\alpha(v)w v^{-1}&=-v w v^{-1} = |v|^{-2} v w v = |v|^{-2}\left( u w^\perp v + v w^\parallel v \right) \\
        			&= |v|^{-2}(-vv w^\perp - \eta(v,w^\perp)v - |v|^2w^\parallel)\\
        			& = w^\perp - w^\parallel = l(v)w.
        		\end{split}
        	\end{equation*}
        \item Setting $\{v_a\}$ as the usual orthonormal basis of $V$, let $S\in\ker(l)$, then for all $u\in V$, $\alpha(S)u S^{-1}= u$, implying $\alpha (S) u = u S$. Splitting $S=S_0 + S_1$ into even and odd part, we obtain
            \begin{equation*}
                u S_0=S_0 u \qquad uS_1=-S_1 u.
            \end{equation*}
        Without loss of generality, we can set $S_0=p_0 + v_1 p_1$, where $p_0$ and $p_1$ are respectively even and odd polynomials in $v_2,\cdots,v_D$. Then, using the above equation with $u=v_1$, we see
            \begin{equation*}
                v_1p_0 + v_1^2 p_1 = p_0 v_1 + v_1 p_1 v_1 = p_0 v_1 - v_1^2 p_1,
            \end{equation*}
        hence $v_1^2p_1=0$, implying $p_1=0$. As a consequence $S_0$ does not contain $v_1$, but this procedure can be iterated for all basis elements $v_a$, hence one must have $S_0=\lambda \mathbb{1}$ for some $\lambda\in \mathbb{R}^*$. The same argument can be repeated for $S_1$, hence showing $\ker{l}=\mathbb{1}\cdot\mathbb{R}^*$.
    \end{enumerate}
\end{proof}

\begin{theorem}
    The following is a short exact sequence
    \begin{equation}
        1\rightarrow \mathbb{R}^* \rightarrow \Gamma(V) \rightarrow O(V) \rightarrow 1
    \end{equation}
\end{theorem}
\begin{proof}
    By point \ref{lem: ker l} of the previous lemma, $\ker(l)=\mathbb{R}^*$, hence we just need to show $l$ is surjective onto $O(V)$. Notice
        \begin{align*}
            \eta\big(l(S)u,l(S)w)&=-\frac{1}{2}\big( l(S)ul(S)w + l(S)wl(S)u \big)\\
            &= -\frac{1}{2}\big( l(S)ul(\alpha(S))w + l(S)w l(\alpha(S))u \big)\\
            &= -\frac{1}{2} \alpha(S) (uw+wu) \alpha(S^{-1})\\
            & = \eta(u,w),
        \end{align*}
    hence proving that $l:\Gamma(V)\rightarrow O(V)$ and $l$ is a homomorphism.

    Now, by Cartan-Dieudonne theorem, for all $R\in O (V)$, $R=R_1\cdots R_k$ for $k\leq D=\mathrm{dim}(V)$ and $R_i$ are reflections. By point \ref{lem: reflections} we know there exist unit vectors $u_i\in V$ such that $R_i=l(u_i)$ and therefore $R=l(u_1)\cdots l(u_k)=l(u_1\cdots u_k)$, hence showing that $l$ is surjective.
        
\end{proof}

One can define the further subgroup $S(V)\subset \mathcal{C}^*(V)\subset \mathcal{C}(V)$ of invertible elements $S$ whose inverse is proportional to their transpose, namely such that $S \overline{S}\propto \mathbb{1}$. 

\begin{definition}[Pin and Spin groups]

    We define the \text{Pin group} Pin$(V)$ to be the subgroup of $S(V)$ generated by unit vectors (i.e. such that $v^2=\eta(v,v)=\pm {1}$), while the \text{Spin group} Spin$(V)$ is defined to be the intersection of Pin$(V)$ with the even Clifford subalgebra $\mathcal{C}(V)$. In other words
        \begin{align}
            \text{Pin}(V)&:=\{u_1\cdots u_k \hspace{1mm}\vert \hspace{1mm} u_i^2=\pm 1\}\\
            \text{Spin}(V)&:=\{u_1\cdots u_k \hspace{1mm}\vert \hspace{1mm} \text{ $k$ even and } u_i^2=\pm 1\}=\text{Pin}(V)\cap \mathcal{C}_0(V).
        \end{align}
\end{definition}

Elements in Spin$(V)$ are products of an even number of unit vectors, $S=u_1 u_2\cdots u_{2k}$. In this case it is easy to find the inverse of $S$, as	
\begin{equation*}
	S^{-1}=\frac{\pm 1}{|u_1 |^2\cdots |u_{2k}|^2}u_{2k}\cdots u_2 u_1
\end{equation*}

As an immediate consequence of the above theorem, we have the following
\begin{corollary}
    The restriction of $l$ to the Pin and Spin groups defines the following short exact sequences 
    \begin{align}
        &\nonumber 1\rightarrow \mathbb{Z}_2 \rightarrow \mathrm{Pin}(V) \rightarrow O(V) \rightarrow 1, \\
        &\label{SES: S}1\rightarrow \mathbb{Z}_2 \rightarrow \mathrm{Spin}(V) \rightarrow SO(V) \rightarrow 1.
    \end{align}
\end{corollary}

\subsubsection{Lie Algebra of Spin group}

\begin{proposition}
    Let $V$ be a $D$--dimensional real vector space. Lie$($Spin$(V))$ is a Lie subalgebra of $\mathcal{C}(V)$, given by
        \begin{equation*}
            \mathrm{Lie(Spin}(V))=\wedge ^2 V
        \end{equation*}
\end{proposition}
\begin{proof}
This can be seen by noticing that the double cover $l\colon \text{Spin}(V)\rightarrow SO(V)$ reduces to an isomorphism of Lie algebras (locally their tangent space at the identity is the same)
    \begin{equation*}
        \begin{split}
            \dot{l}\colon &\mathfrak{spin}(V) \rightarrow \mathfrak{so}(V)\\
            & a \longmapsto \dot{l}(a)=[a,\cdot],
        \end{split}
    \end{equation*}
    where, for all $u\in V$, the $[a,u]\in SO(V)$ is given by
        \begin{equation*}
            [a,u]:=\frac{\partial}{\partial t}\big\vert_{t=0}(e^{-ta}u e^{ta}).
        \end{equation*}

Now, knowing that $\{ v_a \wedge v_b \}$ is a basis for $\mathfrak{so}(V)$, we compute a basis for $\mathfrak{spin}(V)$. 

Define $v_{ab}:=\frac{1}{4}[v_a,v_b]$, then, for all $u=u^c v_c\in V$ 
    \begin{align*}
        \dot{l}(v_{ab})u&=\frac{1}{4}[[v_a,v_b],u]=\frac{1}{2}[v_a v_b, u]\\
        &=\frac{1}{2}\left( v_a v_b u - u v_a v_b \right)= \frac{1}{2}\left( v_a v_b u - u v_a v_b + v_a u v_b - v_a u v_b \right)\\
        &= \eta(u,v_a)v_b - \eta(v_b,u)v_a=u^c (\delta^d_b \eta_{ac} - \delta_a^d \eta_{bc}) v_d,\\
    \end{align*}
    hence  
        \begin{equation*}
            \dot{l}(v_{ab})^d_c = \delta^d_b \eta_{ac} - \delta_a^d \eta_{bc} = -(M_{ab})^d_c
        \end{equation*}
    where $M_{ab}$ are the generators of the Lorentz group $SO(V)$ in the fundamental representation. This implies that $-\frac{1}{4}[v_a,v_b]$ defines a basis for $\mathfrak{spin}(V)$.

\end{proof}

\subsection{Representations}

Given the definition of the Pin and Spin groups seen respectively as subgroups of $\mathcal{C}(V)$ and $\mathcal{C}_0(V)$, classifying irreducible representations of $\mathcal{C}(V)$ and $\mathcal{C}_0(V)$ will automatically produce a classification of irreps of the Pin and Spin groups, which are called respectively pinor and spinor representations.


Looking at table \ref{table: Clif alg}, we can already classify the irreducible pinor representations, as $\mathbb{H}(N)$ and $\mathbb{R}(N)$ have a unique irreducible representation given respectively by $\mathbb{H}^N$ and $\mathbb{R}^N$, whereas $\mathbb{C}(N)$ has two, one isormorphic to $\mathbb{C}^N$ and the complex conjugate one. Therefore the number of irreducible  pinor representations is given by
    \begin{equation*}
        p_{r,s}=\begin{cases}
            2 \text{ if } r-s=1,3 \text{ mod 4},\\
            1 \text{ if } r-s=2,4 \text{ mod 4}.
        \end{cases}
    \end{equation*}
Table \ref{table: Clif alg} also tells us whether the representation is real, complex or quaternionic.

For the spinor representations, since Spin is a subspace of the even Clifford algebra, we only need to look at table \ref{table: Clif alg}, which implies that the number of irreducible inequivalent spinor representations is
    \begin{equation*}
        s_{r,s}=\begin{cases}
            2 \text{ if } r-s=2,4 \text{ mod 4},\\
            1 \text{ if } r-s=1,3 \text{ mod 4}.
        \end{cases}
    \end{equation*}
Notice that in the even dimensional case $D=2k$ there are two inequivalent irreducible spinor representations (known as Weyl representations), which correspond to the Weyl spinors and can be understood by looking at the volume element $v_*=v_1\cdots v_D$. Being $v_*$ the product of an even number of generators, it anticommutes with them, i.e. $\{v_*,v_a\}=0$ for all $a=1,\cdots,D$, but it commutes with all the elements in the even Clifford subalgebra, hence with all the group elements in Spin$(V)$, which implies that it must act as a scalar in the spinor representations. This means that the inequivalent Weyl representations can be labelled by the eigenvalues of $v_*$. Furthermore, a straightforward computation gives, for even $D=r+s=2k$
    \begin{equation}\label{eq: squared v_*}
        v_*^2=(-1)^{\frac{r-s}{2}}\mathbb{1}.
    \end{equation}
One can classify the inequivalent spinor representation as follows
\begin{itemize}
    \item $r-s=0$ mod 8. There are two inequivalent real spinor representations, of real dimension $2^{\frac{D-2}{2}}$, labeled by the eigenvalue of $v_*$ being $1$ or $-1$;
    \item $r-s=1,7$ mod 8. There is a unique spinor representation, which is real and of real dimension $2^{\frac{D-1}{2}} $;
    \item $r-s=2,6$ mod 8. There are two inequivalent complex spinor representations, of complex dimension $2^{\frac{D-2}{2}}$, labeled by the eigenvalue of $v_*$ being $i$ or $-i$;
    \item $r-s=3,5$ mod 8. There is a unique spinor representation, which is quaternionic and of quaternionic dimension $2^{\frac{D-3}{2}}  $;
    \item $r-s=4$ mod 8. There are two inequivalent quaternionic spinor representations, of quaternionic dimension $2^{\frac{D-4}{2}}$, labeled by the eigenvalue of $v_*$ being $1$ or $-1$;
\end{itemize}

\subsubsection{Complex representations and the Lorentzian signature case}

As it is significantly easier to deal with complex Clifford algebras, we turn our attention to complex representations of $\mathcal{C}(D)$. 

We recall
    \begin{equation}
        \mathcal{C}(2k)\simeq \mathbb{C}(2^k) \qquad \text{and}\qquad \mathcal{C}(2k+1)\simeq \mathbb{C}(2^k)\oplus\mathbb{C}(2^k),
    \end{equation}
which implies there are faithful representations
    \begin{align}
        &\label{eq: pinor cplx rep even}\Gamma_{(2k)}:\mathcal{C}(2k)\rightarrow \mathrm{End}(\mathbb{C}^{2^k})\\
        &\label{eq: pinor cplx rep odd}\Gamma_{(2k+1)}:\mathcal{C}(2k+1)\rightarrow \mathrm{End}(\mathbb{C}^{2^k})\oplus\mathrm{End}(\mathbb{C}^{2^k}),
    \end{align}
where $\Gamma_{2k}$ is irreducible and $\Gamma_{2k+1}$ splits into two irreducible representations. These are precisely the pinor representations of the complex Clifford algebra. 

\begin{remark}
    The above irreducible representations are unique up to conjugacy with unitary matrices. From now on, we drop the subscript and denote such representations just by $\Gamma$. 
\end{remark}

\begin{proposition}\label{prop: cmplx clif lorentz gamma}
    Let $s=1$ (i.e. only one time--like direction), $d:=r$ and $k:=\lfloor \frac{d+1}{2}\rfloor$. Furthermore, let $\{ v_0,v_1\cdots,v_d \}$ be a basis of $V_\mathbb{C}$ such that in the Clifford algebra $v_0^2=\mathbb{1}$ and $v_i v_j + v_j v_i = - 2 \delta_{ij}\mathbb{1}$ for all $i,j=1,\cdots,d$. Then there exists a choice of complex representation $\Gamma$ of $\mathcal{C}(d+1)$ on $\mathbb{C}^{2^k}$ (called Gamma representation) such that
        \begin{enumerate}[(i)]
            \item $\Gamma_0:=\Gamma(v_0)$ is hermitian;
            \item $\Gamma_i:=\Gamma(v_i)$ is anti--hermitian for all $i=1,\cdots,d$;
            \item $\Gamma_0$ defines a hermitian form\footnote{A hermitian form on $V_\mathbb{C}$ is given by an $\mathbb{R}-$bilinear form $\langle-,-\rangle:V\times V \rightarrow \mathbb{C}$ such that for all $v_1,v_2\in V$ and $\lambda \in \mathbb{C}$
    \begin{itemize}
        \item $\langle v_1,\lambda v_2\rangle=\lambda \langle v_1,v_2\rangle$;
        \item $\langle v_1,v_2\rangle^*=\langle v_2,v_1\rangle $, where $(-)^*$ denotes complex conjugation
    \end{itemize}} for all $\psi_1,\psi_2 \in \mathbb{C}^{2^k} $as 
    \begin{equation}
        \langle \psi_1 , \psi_2 \rangle := \psi_1^\dag \Gamma_0 \psi_2,
    \end{equation}
    where $\psi^\dag:=(\psi^*)^t$ denotes the canonical hermitian conjugate in $\mathbb{C}^{2^k}$. Such pairing is called Dirac pairing and, upon defining the Dirac conjugate as $\bar{\psi}:=\psi^\dag \Gamma_0$, can be redefined as 
    \begin{equation*}
        \langle \psi_1 , \psi_2 \rangle = \bar{\psi_1} \psi_2.
    \end{equation*}  
        \end{enumerate}   
\end{proposition}

\begin{remark}
    In the physics context, $\psi$ in $\mathbb{C}^{2^k}$ is called a Dirac spinor, although strictly speaking it is a pinor, since $\mathbb{C}^{2^k}$ is the complex pinor representation as seen in \eqref{eq: pinor cplx rep even} and \eqref{eq: pinor cplx rep odd}. Furthermore, following Dirac's nomenclature, the matrices $\Gamma_a$ in $\mathbb{C}(2^k)$ are called gamma matrices.
\end{remark}

\begin{remark}\label{rem: invariance under spin of dirac pairing}
    The Dirac conjugate definition extends to any operator $A\in$End$(\mathbb{C}^{2^k})$ as $$\bar{A}:=\Gamma_0^{-1}A^{\dagger}\Gamma_0.$$  
    As it turns out, from the above proposition it follows
        \begin{equation}
            \Gamma_a^\dag := \Gamma_0^{-1} \Gamma_a \Gamma_0 \qquad \forall a=0,\cdots,d,
        \end{equation}
    and noticing that $\Gamma_0^{-1}=\Gamma_0 $, it is easy to see that the gamma matrices are invariant under Dirac conjugation, i.e. $\bar{\Gamma}_a=\Gamma_a$.
    
    Furthermore, it is possible to prove that the spin group representation on $\mathbb{C}^{2^k} $ induced by the gamma representation is unitary. Indeed, recalling that $\{\frac{1}{4}v_{ab}\}$ defines a basis of the Lie algebra $\mathfrak{spin}(1,d)$, and having defined $\Gamma_{ab}:=\Gamma(v_{ab})=\frac{1}{2}\Gamma(v_a v_b - v_b v_a)=\frac{1}{2}(\Gamma_a\Gamma_b - \Gamma_b\Gamma_a ) $, one sees 
        \begin{align*}
            \Gamma_0^{-1}\left( \Gamma_{a}\Gamma_b \right)^\dag\Gamma_0&=\Gamma_0^{-1}\Gamma_b^\dag \Gamma_a^\dag \Gamma_0 =\Gamma_0^{-1}\Gamma_0^{-1}\Gamma_b \Gamma_0 \Gamma_0^{-1} \Gamma_a \Gamma_0=\Gamma_b \Gamma_a,
        \end{align*}
    therefore $\Gamma_0^{-1}(\Gamma_{ab})^\dag\Gamma_0=\Gamma_{ba}=-\Gamma_{ab} $, hence implying (expanding the exponential)   
        \begin{align}  
            \Gamma_0^{-1}\mathrm{exp}\left( \frac{1}{4}\omega^{ab}\Gamma_{ab} \right)^\dag\Gamma_0  =\mathrm{exp}\left( -\frac{1}{4}\omega^{ab}\Gamma_{ab} \right)=\mathrm{exp}\left( \frac{1}{4}\omega^{ab}\Gamma_{ab} \right)^{-1}. 
        \end{align}
      
\end{remark}

\begin{proof}
    We start by the case of $D=3+1$ and prove the proposition by induction\footnote{The cases where $D=1$ and $D=2$ have appeared in previous examples, while the case for $D=3$ can be derived from the $D=2$ using the same induction method}. Consider the Pauli matrices $\sigma_a$ defined by
        \begin{align*}
            &\sigma_0= \begin{pmatrix}
                1 & 0\\
                0 & 1
            \end{pmatrix}, \quad\sigma_1= \begin{pmatrix}
                0 & 1\\
                1 & 0
            \end{pmatrix},\quad \sigma_2= \begin{pmatrix}
                0 & i\\
                -i & 0
            \end{pmatrix}, \quad\sigma_3= \begin{pmatrix}
                1 & 0\\
                0 & -1
            \end{pmatrix},
        \end{align*}
    and set $\bar{\sigma}_0=\sigma_0$ and $\bar{\sigma}_i=-\sigma_i$ for $i=1,2,3.$ Then a choice of gamma matrices is given by
        \begin{equation*}
        \Gamma_a=\begin{pmatrix}
            0 & \sigma_a \\
            \bar{\sigma}_a & 0
        \end{pmatrix}.
        \end{equation*}
    One can easily check that this choice satisfies the Clifford condition, while $\Gamma_0$ is hermitian and $\Gamma_i$ antihermitian for $i=1,2,3.$    

    The fact that $\langle-,-\rangle$ is a hermitian form is an immediate consequence of $\Gamma_0$ being hermitian.

    Now, assuming there exists a gamma representation for $D=2k$, given by matrices $\Gamma_a$, one can define a gamma matrices $\Gamma'_a$ for $a=0,\cdots,D$ (i.e. a gamma representation for $D+1$) as follows 
        \begin{equation*}
            \Gamma'_a:=\begin{cases}
                \Gamma_a \quad \text{for $a\leq d$}\\
                \Gamma'_{d+1}=\alpha\Gamma_*=\alpha\Gamma_0\Gamma_1\cdots\Gamma_{d}
            \end{cases}    
        \end{equation*}
    where $$ \alpha=\begin{cases}
        1 \quad\text{ for $k$ even}\\ 
        i \quad \text{ for $k$ odd}
        \end{cases} $$    

    In a similar way, starting from gamma matrices for $D=2k$, one obtains gamma matrices $\Gamma_a''$ representing the complex $D+2$ Clifford algebra as
        \begin{equation*}
            \Gamma''_{a\leq d}:=\begin{pmatrix}
                0 & \Gamma_a \\
                \Gamma_a & 0
            \end{pmatrix}, \qquad \Gamma''_{d+1}:=\begin{pmatrix}
                0 & \mathbb{1}\\
                -\mathbb{1} & 0
            \end{pmatrix}, \qquad \Gamma''_{d+2}=\begin{pmatrix}
                i\mathbb{1} & 0\\
                0 & i\mathbb{1}
            \end{pmatrix}.
        \end{equation*}
\end{proof}

\subsection{Charge conjugation and Majorana spinors in the Lorentzian signature}

Before giving the definition of Majorana spinors, we first notice that the sets $\{ \pm\Gamma_a^* \}$ define two new (equivalent) representations of the complex Clifford algebra $\mathcal{C}(D)$, therefore there must exist a unitary matrix $B$ such that 
    \begin{equation}\label{eq: def B}
        \Gamma_a=\eta B^{-1} \Gamma_a^* B,
    \end{equation}
with $\eta=\pm 1$. By separately taking the complex conjugate and inverting the equation above we find
    \begin{equation*}
        \Gamma_a^*=\eta B^*\Gamma_a (B^*)^{-1} =\eta B\Gamma_a B^{-1},
    \end{equation*}
implying $\Gamma_a=B^{-1}B^* \Gamma_a (B^*)^{-1}B $, which yelds
    \begin{equation*}
        B^*  = \epsilon B^{-1}, \qquad \epsilon=\pm 1.
    \end{equation*}
Notice that since $B$ is unitary, then $B^\dag=B^{-1}=\epsilon B^*$, implying $B^t=\epsilon B$.
In general, $\epsilon$ depends on $\eta$ and can be found using a method due to Scherk \cite{Scherk:1978fh,KUGO1983357}. Upon defining the charge conjugation matrix as
    \begin{equation}
        C:=B^t \Gamma_0,
    \end{equation}
from remark \ref{rem: invariance under spin of dirac pairing} one can see $\Gamma_a^\dag=\Gamma_0 \Gamma_a\Gamma_0$, but at the same time $\Gamma_a^\dag=(\Gamma_a^t)^*=\eta(B^{-1})\Gamma_a^tB^t $, hence finding
    \begin{equation}\label{eq: antisym charge gamma}
        \Gamma_a^t C=\eta C \Gamma_a, \qquad \qquad CC^\dag=\mathbb{1} \qquad \text{and} \qquad C^t=\epsilon \eta  C.
    \end{equation}

Now, first considering $D=2k$, it is clear that the set $\{\Gamma_A\}:=\{\mathbb{1},\Gamma_a,\Gamma_{ab},\cdots,\Gamma_0\Gamma_1\cdots\Gamma_d\}  $, generates the whole algebra of $2^k \times 2^k$ complex matrices, as it is the image of \eqref{Clifford basis} under the gamma representation.
Clearly, for all $A$, $C\Gamma_A$ are still generators of the whole algebra and either symmetric or antisymmetric, depending on $\eta$ as can be seen from \eqref{eq: antisym charge gamma}. The problem of counting how many of these matrices are antisymmetric is addressed in \cite{Scherk:1978fh}, and it depends on $\eta, \epsilon$ and $D$. However, we know that on $\mathbb{C}^{2^k}$ there are $\frac{2^k}{2}(2^k-1)$ independent antisymmetric matrices. Eventually, one finds
    \begin{equation*}
        \epsilon=\mathrm{cos}\left(\frac{\pi}{4}(d-1)\right)-\eta\mathrm{sin}\left(\frac{\pi}{4}(d-1)\right).
    \end{equation*}
In the even-dimensional case one can choose either signs for $\eta=\pm1$, while in the case where $D=2k+1$, one needs to require that $\Gamma_{d+1}$ transforms correctly under $B$ (i.e. as in \eqref{eq: def B}), which fixes $\eta$ as
    \begin{equation*}
        \eta=(-1)^k.
    \end{equation*}

    We can now start discussing about Majorana spinors. 
    \begin{definition}
    A Majorana representation is a particular real representation (of $\mathcal{C}(V)$). It is possible to understand what types of Clifford algebras allow for such real representations by looking at table \ref{table: Clif alg}, but, in the context described above, we regard a Majorana representation as a complex representation endowed with a real structure\footnote{Given a complex linear representation of a Lie group $\rho:G\rightarrow $End$(W)$ on a complex vector space $W$, a real or quaternionic structure is a real linear map $\varphi:W\rightarrow W$ such that 
            \begin{itemize}
                \item $\varphi^2= id_W$ 
                \item $\varphi(\lambda v)=\lambda^* v$, i.e. $\varphi$ is conjugate linear;
                \item $\varphi$ is invariant under $\rho$, i.e. it commutes with the image of all elements of $G$ under $\rho$, $[\varphi,\rho(g)]=0$.
            \end{itemize}}.        
    \end{definition}

        The following theorem allows us to relate the real structure to the charge conjugation matrix.
        \begin{theorem}
            Let $D$ be such that $\epsilon=1$ as defined above, then $$\phi:\mathbb{C}^{2^k}\rightarrow\mathbb{C}^{2^k}\colon \psi \mapsto B \psi^* $$ defines a real structure.
        \end{theorem} 
        \begin{remark}
            In this particular case, one can use the charge conjugation matrix to define a Spin$(d,1)$--invariant complex bilinear form $C:\mathbb{C}^{2^k}\times\mathbb{C}^{2^k}\rightarrow  $ as
            \begin{equation*}
                C(\psi_1,\psi_2):= \psi_1^t \cdot C \cdot \psi_2, \qquad \forall\psi_1,\psi_2\in\mathbb{C}^{2^k}.
            \end{equation*}
            Furthermore, there exist a choice of gamma matrices for which $C$ is real.
        \end{remark}
        \begin{proof}
        Clearly $\phi$ is connjugate linear, while 
            \begin{align*}
                \phi^2 \psi = B B^* \psi = \epsilon \psi = \psi.
            \end{align*}
        
        Lastly, Spin$(d,1)$--invariance amounts to checking that $C$ is Spin$(d,1)$--invariant, namely
            \begin{align*}
                (\Gamma_a \Gamma_b)^t C = \Gamma_b^t \Gamma_a^t C = \eta \Gamma_b^t C \Gamma_a = C \Gamma_b \Gamma_a,
            \end{align*}
        implying $(\Gamma_{ab})^tC=-C\Gamma_{ab} $, hence satisfying 
            \begin{equation*}
                \mathrm{exp}\left( \frac{1}{4}\omega^{ab}\Gamma_{ab} \right)^t C \mathrm{exp}\left( \frac{1}{4}\omega^{ab}\Gamma_{ab} \right) = C.
            \end{equation*}

        \end{proof}
\begin{remark}
    When $\epsilon=-1$, the above proof still holds, but in this case $\phi^2=-id$, hence defining a quaternionic structure.
\end{remark}        

        \begin{definition}
            Assume $\epsilon=1$ for some $k$, then a (s)pinor $\psi\in\mathbb{C}^{2^k} $ satisfying the reality condition
                \begin{equation}
                    \phi(\psi)=B\psi^*=\psi,
                \end{equation}
           is said to be Majorana when $\eta=-1$ and pseudo-Majorana when $\eta=1$.    
        \end{definition}
        \begin{remark}
            It is customary to rephrase the above condition in terms of the charge conjugation matrix, noticing that $B=\epsilon C \Gamma_0$, one obtains that, under the above assumptions, $\psi$ is Majorana when
                \begin{equation*}
                    C \Gamma_0 \psi^*=\psi.
                \end{equation*}
        \end{remark}
        The following table contains information on the allowed values of $\epsilon$ and $\eta$ in various dimensions.
    \begin{table}[h]
    \centering
    \begin{tabular}{|l|l|l|}
    \hline
     & $\eta=1$ & $\eta=-1$ \\ \hline
    $\epsilon=1$  &       $D=1,2,8$ mod $8$       & $D=2,3,4$ mod $8$     \\ \hline
    $\epsilon=-1$ &      $D=4,5,6$ mod $8$      &    $D=6,7,8$ mod $8$     \\ \hline
    \end{tabular}
    \end{table}        
           
\section{Spin coframe formalism, i.e. defining spinor fields on manifolds}\label{sec: spinor fields on manifolds}
In the previous section, we saw the algebraic construction and classification of Clifford algebras and spinors. This section is dedicated to investigating the local structure of such objects in the context of differential geometry, with the goal of providing a framework that allows to treat the definition of supergravity in the same formulation found in \cite{CS2017, CCS2020}. The main part follows \cite{spingeom}, \cite{fatibene2018} and \cite{FatiNoris22} for the spin frame definition. For a detailed review of the "bosonic" coframe formalism, reference \cite{Tec20} is recommended.

\subsection{Basic notions on principal bundles}
In the following, we assume $M$ to be a pseudo-riemannian manifold of dimension $D$.
\begin{definition}
    Let $G$ be a Lie group. A principal $G$-bundle $\pi\colon P\rightarrow M$ is a fiber bundle such that
    \begin{itemize}
        \item There exists a smooth right action $R\colon P\times G \rightarrow P$ which is free, i.e. such that $R(p,e):=p\dot e =p$ for all $p\in P$, letting $e\in G$ be the identity;
        \item $\pi\colon P \rightarrow M $ is diffeomorphic as a bundle to $ P\rightarrow P/G$.
    \end{itemize}
\end{definition}
\begin{remark}
    Notice that, since $R$ is free, any orbit $\mathcal{O}_p:=\{ q\in P \hspace{1mm}|\hspace{1mm} \exists g\in G \text{ s.t. } q=p\cdot g  \}=[p]\in P/G$ is isomorphic to $G$. Then points $x\in M$ is in one-to-one correspondence with orbits $[p]\in P/G$, and each fiber is isomorphic to the group $G$, as $\pi^{-1}(x)=[p]=\mathcal{O}_p\simeq G $.
\end{remark}
\begin{definition}
    Given a $G$-principal bundle $P$, a trivialization of $P$ is a collection $(U_\alpha,\phi_\alpha)$, with $\alpha$ is an element of an index set $I$, such that
    \begin{itemize}
        \item $\mathcal{U}:=\{ U_\alpha \}_{\alpha\in I} $ is an open cover of $M$,
        \item $\phi_\alpha\colon\pi^{-1}(U_\alpha)\rightarrow U_\alpha \times G $ are diffeomorphisms
        \item letting $U_{\alpha\beta}:= U_\alpha \cap U_\beta $, transition functions $\phi_{\alpha\beta}\colon U_{\alpha\beta}\times G \rightarrow U_{\alpha\beta}\times G$ are given by (smooth) functions $g_{\alpha\beta} \colon U_{\alpha\beta}\rightarrow G $ via left action as $\phi_{\alpha\beta}\colon (x, h)\mapsto (x,g_{\alpha\beta}\cdot h ) $ and must respect the cocycle identity 
            \begin{equation}
                g_{\alpha \beta}\cdot g_{\beta \gamma}\cdot g_{\gamma \alpha}= e, \qquad \text{for all}\hspace{1mm} x\in U_{\alpha\beta\gamma}.
            \end{equation}
    \end{itemize}
\end{definition}

\begin{remark}
    In general, a principal bundle can be recovered by pasting together its local data, i.e. pasting the local products $\{ U_\alpha\times G\} $ via the transition functions $g_{\alpha\beta} $. Indeed, it is possible to show that any principal $G-$bundle $P$ is equivalent to a pair $(\mathcal{U}, \{ g_{\alpha\beta} \} )$, where $\mathcal{U}$ is an open cover and $\{ g_{\alpha\beta}\colon U_{\alpha\beta}\rightarrow G \} $ are functions satisfying the cocycle condition.\footnote{The cocycle condition is equivalent to the Cech coboundary condition, and $g_{\alpha \beta}$ is nothing but a Cech 1--cocycle with coefficients in $G$ (to be precise, with coefficients in the sheaf of germs of smooth maps to $G$).}
\end{remark}

\begin{definition}
    Two $G$-principal bundles $P$ and $P'$ over $M$ are equivalent if there exists a homeomorphism $H:P\rightarrow P'$ such that the following diagram commutes
    \begin{equation*}
    \begin{tikzcd}
    	P && {P'} \\
    	& M
    	\arrow["H", from=1-1, to=1-3]
    	\arrow["\pi"', from=1-1, to=2-2]
    	\arrow["\pi'", from=1-3, to=2-2]
    \end{tikzcd}
    \end{equation*} 
    and such that $H$ is equivariant, i.e. $H(p\cdot g)=H(p)\cdot g$ for all $g\in G$ and $p\in P$.
\end{definition}

It is interesting to understand such definition at the level of local trivialization, which will allow us to describe the set of inequivalent principal $G$--bundles over $M$. First of all, let $P$ and $P'$ be defined respectively by $(\mathcal{U}, \{ g_{\alpha\beta} \} )$ and $(\mathcal{U}, \{ g'_{\alpha\beta} \} )$ as in the above remark. They induce trivializations $(\phi_\alpha)$ and $(\phi'_\alpha)$, which allow to define 
    \begin{equation*}
        H_\alpha:= \phi'_\alpha \circ H \circ \phi_\alpha^{-1} \colon U_\alpha \times G \rightarrow U_\alpha \times G.
    \end{equation*}
Now, since $\pi'\circ H=\pi$, we must have that $H_\alpha(x,g)=(x,h_\alpha(x,g))$ for some $h_\alpha\colon U_\alpha \times G \rightarrow G$. Now, using equivariance, we obtain 
    \begin{equation*}
        H_\alpha(x,g\cdot f)=H_\alpha(x,g)\cdot f \qquad \Rightarrow \qquad h_\alpha(x,g\cdot f)= h_\alpha(x,g)\cdot f,
    \end{equation*}
which implies that $H_\alpha(x,g)=(x,h_\alpha(x,e)\cdot g )=(x,g_\alpha(x)\cdot g)$, having defined $g_\alpha(x):=h_\alpha(x,e)$. The relation between the transition functions can be understood by noticing that, by definition of equivalence, the following diagram must commute
    \begin{equation*}
\begin{tikzcd}
	{U_{\alpha\beta}\times G } && {\pi^{-1}(U_{\alpha\beta})} && {U_{\alpha\beta}\times G } \\
	{U_{\alpha\beta}\times G } && {(\pi')^{-1}(U_{\alpha\beta})} && {U_{\alpha\beta}\times G }
	\arrow["{\phi_\alpha^{-1}}", from=1-1, to=1-3]
	\arrow["{\phi_{\beta\alpha}}", curve={height=-24pt}, from=1-1, to=1-5]
	\arrow["{H_\alpha}", from=1-1, to=2-1]
	\arrow["{\phi_\beta}", from=1-3, to=1-5]
	\arrow["H", from=1-3, to=2-3]
	\arrow["{H_\beta}", from=1-5, to=2-5]
	\arrow["{\phi_\alpha'^{-1}}"', from=2-1, to=2-3]
	\arrow["{\phi_{\beta\alpha}'}"', curve={height=24pt}, from=2-1, to=2-5]
	\arrow["{\phi_\beta'}"', from=2-3, to=2-5]
\end{tikzcd} 
    \end{equation*}
therefore $\phi_{\beta\alpha}'= H_\beta \circ \phi_{\beta\alpha} \circ H_\alpha^{-1} $, implying
    \begin{equation*}
        g'_{\alpha\beta }= g_\alpha^{-1} \cdot g_{\alpha\beta} \cdot g_\beta.
    \end{equation*}
Hence $(\mathcal{U}, \{ g_{\alpha\beta} \} )$ and $(\mathcal{U}, \{ g'_{\alpha\beta} \} )$ define equivalent bundles $P$ and $P'$ iff there exists a family $g_\alpha:U_\alpha\rightarrow G$ of smooth functions such that $g'_{\alpha\beta }= g_\alpha^{-1} \cdot g_{\alpha\beta} \cdot g_\beta$. Upon inspection, one realizes that this is nothing but a Cech-coboundary condition (in the multiplicative sense), and therefore $g'_{\alpha\beta }$ and $g_{\alpha\beta}$ only "differ by an exact term", where $g_\alpha$ acts as a "Cech 0-cochain". Therefore one can see an equivalence class of principal $G$--bundles as an element of $H^1(\mathcal{U};G).$

Letting $(\mathcal{U}_i)$ be a family of open covers such that for $i>j$ $\mathcal{U}_i\subset \mathcal{U}_j$, then one can define $H^1(M;G)$ as the direct limit (in the categorical sense)
    \begin{equation*}
        H^1(M;G)=\textbf{lim}_i H^1(\mathcal{U}_i;G).
    \end{equation*}
 Notice that this set is strictly speaking not a group, but contains an identity given by the trivial principal bundle $M\times G$. If $G$ is abelian, then $H^1(M;G)$ is just the first Cech cohomology group with coefficients in $G$.

\begin{definition}
    The frame bundle $LM\rightarrow M$ is a principal GL$(D,\mathbb{R})$--bundle defined by
        \begin{equation*}
            LM=\bigcup_x L_xM, \qquad L_x M :=\{ e_a=(e_0,\cdots,e_d)\hspace{1mm}\vert\hspace{1mm} (e_a) \text{ is a basis of }T_xM\}.
        \end{equation*}
    with trivialisation given by $\phi_\alpha\colon \pi^{-1}(U_\alpha)\rightarrow U_\alpha \times \mathrm{GL}(D,\mathbb{R})\colon (x,e_a) \mapsto (x, e_a^\mu) $, having set $\mu=0,\cdots,d$    
\end{definition}
\begin{remark}
    One can see that the transition functions between two charts with local coordinates $\{x\}$ and $\{x'\}$ act via left action as ${e'}_a^\mu =  \frac{\partial x'^\mu}{\partial x^\nu} e_a^\nu $  .
\end{remark} 

Assuming that $M$ is orientable, and having chosen a Lorentzian metric $g$ on it, one can define the orthonormal frame bundle as the subbundle of the frame bundle containing orthonormal frames, i.e.
    \begin{equation*}
        SO(M,g):=\{ e_a \in LM \hspace{1mm}|\hspace{1mm} g(e_a,e_b)=\eta_{ab} \},
    \end{equation*}
where $\eta_{ab}$ is the Minkowski metric.     

\begin{remark}\label{rem: ON frame}
   Notice that  $SO(M,g)$ is a principal SO$(d,1)$--bundle. Furthermore, for a given metric $g$, there exist more than one ON basis, as for any $e_a$ ON and for any $\Lambda\in$ SO$(d,1)$, also $e'_b = e_a \Lambda ^a_b$ satisfies $g(e'_a,e'_b)=\eta_{ab}$. However, the viceversa is not true, indeed for each ON basis $e_a$ there is a unique metric $g$ with respect to which it is orthonormal.
\end{remark}

As it turns out, it is particularly useful to consider the dual notion of an ON frame, namely an ON coframe, the dual basis $e^b$ with respect to a given frame $e_a$, i.e. such that
    \begin{equation*}
        e^b(e_a)=\delta^b_a.
    \end{equation*}
This motivates the following definition
\begin{definition}
    Given a principal SO$(d,1)$--bundle $P_{SO}$, a veilbein map is a principal bundle morphism $\tilde{e}\colon P_{SO}\rightarrow LM$ satisfying verticality and equivariance, i.e. such that the following two diagrams commute 
    \begin{equation*}
\begin{tikzcd}
	{P_{SO}} && LM && {P_{SO}} & LM \\
	& M &&& {P_{SO}} & LM
	\arrow["{\tilde{e}}", from=1-1, to=1-3]
	\arrow["{\pi}", from=1-1, to=2-2]
	\arrow["{\pi}"', from=1-3, to=2-2]
	\arrow["{\tilde{e}}", from=1-5, to=1-6]
	\arrow["{\cdot \Lambda}", from=1-5, to=2-5]
	\arrow["{\cdot \Lambda}", from=1-6, to=2-6]
	\arrow["{\tilde{e}}", from=2-5, to=2-6]
\end{tikzcd}
    \end{equation*}
where $\Lambda$ is an element of SO$(d,1)$ possibly seen as an element of GL$(d+1,\mathbb{R}).$    
\end{definition}

Choosing a local section $s_\alpha:U_\alpha \rightarrow P_{SO} $, on the overlap of two patches $U_{\alpha\beta}$ transition functions $\Lambda_{\alpha\beta}\colon U_{\alpha\beta}\rightarrow \mathrm{SO}(d,1) $ act via right action as
    \begin{equation*}
        s_\beta = s_\alpha \cdot \Lambda_{\alpha\beta}.
    \end{equation*}
As it turns out, it is sufficient to know $\tilde{e}$ on $s_\alpha$ to know it on the whole $\pi^{-1}(U_\alpha) $, indeed $\tilde{e}(s_\alpha(x))=(x, {}_\alpha e_a(x)) $, where $e_a(x)$ is a frame defining a basis of $T_xM$, then thanks to equivariance, for all $p\in\pi^{-1}(U_\alpha) $, there exists a $\Lambda\in\mathrm{SO}(d,1)$ such that $p=s_\alpha \cdot \Lambda $, hence $\tilde{e}(p)=\tilde{e}(s_\alpha\cdot \Lambda)= \tilde{e}(s_\alpha)\cdot \Lambda$.

It is then clear that a vielbein map uniquely defines a family of frames differing by orthonormal transformations on overlaps of the patches. It is also easy to see that the viceversa is true, and as a consequence, keeping in mind remark \ref{rem: ON frame}, a vielbein map uniquely defines a metric $g$ (with respect to which $e_a$ is ON) on $M$ via
the dual frame, i.e.
    \begin{equation*}
        g_{\mu\nu}=e^a_\mu e^b _\nu \eta_{ab}.
    \end{equation*}
It becomes even clearer when one takes the image of $P_{SO}$ under $\tilde{e}$, finding
    \begin{equation*}
        \tilde{e}(P_{SO})=\{ (x,e_a)\hspace{1mm}\vert \hspace{1mm} (e_a) \text{ is a basis of $T_xM$, and } g(e_a,e_b)=\eta_{ab}\}\simeq SO(M,g).
    \end{equation*}
This observation motivates the following definition:
\begin{definition}
    Let $(V,\eta)$ be a real $D-$dimensional vector space endowed with the Minkowski metric, and let $\rho: \mathrm{SO}(d,1)\rightarrow V$ be the fundamental representation of the Lorentz group on $V$, then the Minkwoski bundle $\mathcal{V}$ is the associated bundle\footnote{Here $P_{SO}\times_\rho V$ is defined to be the quotient $P_{SO}\times V/\sim$, where $(p,v)\sim (q,w)$ if there exists a $\Lambda\in \mathrm{SO}(d,1)$ such that $q=p\cdot \Lambda$ and $w=\rho(\Lambda)^{-1}\cdot v$}
        \begin{equation*}
            \mathcal{V}=P_{SO}\times_\rho V,
        \end{equation*}    
\end{definition}
With this definition, it is clear that the vielbein map is in 1-to-1 correspondence with linear isomorphisms between $TM$ and $\mathcal{V}$, as they are given by coframes (called vielbein field) dual to the ones defined by the vielbein map. In particular, choosing a local basis $\{v_a\}$ of $\mathcal{V}$ and local coordinates $x$ on $M$, one has
    \begin{equation*}
        e\colon TM \overset{\sim}{\rightarrow } \mathcal{V}, \qquad e=e_\mu^a dx^\mu v_a \quad \text{s.t.}\quad e_\mu^a e^\mu_b=\delta^a_b.
    \end{equation*}

\subsection{Spin structures and the equivalence with spin (co)frames}
    
\begin{definition}
    Let $P_{s}$ be a Spin$(d,1)$--principal bundle over $(M,g)$, a spin structure is a pair $(P_{s},\Sigma)$ such that $\Sigma\colon P_s\rightarrow SO(M,g)$ is an equivariant principal morphism, i.e. such that the following diagrams commute
    \begin{equation*}
    \begin{tikzcd}
    	{P_{s}} && SO(M,g) && {P_s} & SO(M,g) \\
    	& M &&& {P_{s}} & SO(M,g)
    	\arrow["{\Lambda}", from=1-1, to=1-3]
    	\arrow["{\pi}", from=1-1, to=2-2]
    	\arrow["{\pi}"', from=1-3, to=2-2]
    	\arrow["{\Lambda}", from=1-5, to=1-6]
    	\arrow["{\cdot S}", from=1-5, to=2-5]
    	\arrow["{\cdot l(S)}", from=1-6, to=2-6]
    	\arrow["{\Lambda}", from=2-5, to=2-6]
    \end{tikzcd}
    \end{equation*}
where $S\in\mathrm{Spin}(d,1)$ and $l:\mathrm{Spin}(d,1)\rightarrow \mathrm{SO}(d,1)$ is the double covering defined in the previous chapter.
\end{definition}
\begin{remark}
    In general it is not true that every orientable pseudoriemannian manifold admits a spin structure, but, as we will see, there are topological requirements that need to be assumed for it to be true.
\end{remark}

We notice that the notion of spin structure is similar to the one of equivalence of principal bundles, so it might be useful to rephrase the problem of understanding when a spin structure exists in terms of equivalence of bundles.

We saw earlier that $H^1(M;G)$ is the set of inequivalent principal $G$--bundles. Borrowing some results from the theory of Cech cohomology, one can prove that if 
\begin{equation*}
    1\rightarrow K \overset{i}{\rightarrow} G \overset{j}{\rightarrow} G' \rightarrow 1
\end{equation*}
is a short exact sequence of topological groups, then there is an exact sequence at the level of cohomology, given by
    \begin{equation*}
        1\rightarrow H^0(M;K) \overset{i_*}{\rightarrow} H^0(M;G) \overset{j_*}{\rightarrow} H^0(M;G') \overset{\partial_*}{\rightarrow} H^1(M;K) \overset{i_*}{\rightarrow} H^1(M;G) \overset{j_*}{\rightarrow}H^1(M;G'),
    \end{equation*}
where $\partial$ is the Cech coboundary operator and $H^0(M;G)$ is the global sections of $G$ seen as 0-cocycles.\footnote{Indeed the cocycle identity is exactly the requirement that the local sections can be glued to a global one on the overlap of the patches $U_\alpha$.}
It is also possible to prove, if $K$ is abelian, that the sequence can be extended to
    \begin{equation*}
        \cdots \rightarrow H^1(M;K) \rightarrow H^1(M;G) \rightarrow H^1(M;G')\rightarrow H^2(M;K).
    \end{equation*}

Therefore, considering the short exact sequence \eqref{SES: S} $0\rightarrow \mathbb{Z}_2 \rightarrow \mathrm{Spin}(d,1) \overset{l}{\rightarrow} \mathrm{SO}(d,1)\rightarrow 0$, we can define the second Stiefel-Whitney class as the induced map $w_2$ in the exact sequence
    \begin{align*}
        & w_2:H^1(M;\mathrm{SO}(d,1))\rightarrow H^2(M;\mathbb{Z}_2)\\
        & H^1(M;\mathrm{Spin}(d,1))\overset{l_*}{\longrightarrow}(M;\mathrm{SO}(d,1)) \overset{w_2}{\longrightarrow} H^2(M;\mathbb{Z}_2)
    \end{align*}

\begin{theorem}
    $(M,G)$ admits a spin structure if and only if $w_2([SO(M,g)])=0$.
\end{theorem}    
\begin{proof}
    When considering the orthonormal frame bundle $SO(M,g)$, and in particular its equivalence class $[SO(M,g)]\in H^1(M;\mathrm{SO}(d,1))$, we see that the second Stiefel-Whitney class of $[SO(M,g)]$ vanishes if and only if $[SO(M,g)]\in \mathrm{Im}(l_*)$, which tells us that the orthonormal bundle is induced by a Spin bundle, and in particular $l_*$ defines a spin structure. 

    To see it more explicitly, let $ (\mathcal{U},\{g_{\alpha\beta}\}) $ be a cocycle representing $SO(M,g)$, with $\mathcal{U}$ defined such that each non empty $U_{\alpha\beta} $ is simply connected. We can lift the $g_{\alpha\beta} $ to functions $\{ \tilde{g}_{\alpha\beta}\colon U_{\alpha\beta}\rightarrow \mathrm{Spin}(d,1) \}$ and define $K_{\alpha\beta\gamma}:= \tilde{g}_{\beta\gamma}\cdot (\tilde{g}_{\alpha\gamma})^{-1}\cdot \tilde{g}_{\alpha\beta} $\footnote{Notice that this is almost a coboundary as $ \tilde{g}_{\beta\gamma}(\tilde{g}_{\alpha\gamma})^{-1}\tilde{g}_{\alpha\beta} = (\partial \tilde{g})_{\alpha\beta\gamma}$, however it is not because $\tilde{g}_{\alpha\beta}$ does not take values in $\mathbb{Z}_2$, namely $\tilde{g}_{\alpha\beta}$ is not a $\mathbb{Z}_2$-cochain.} on $U_{\alpha\beta\gamma}$. Clearly $l(K_{\alpha\beta\gamma})=1$ as this is exaclty the cocycle identity for $SO(M,g)$, implying $K_{\alpha\beta\gamma}\in\mathbb{Z}_2 $. Furthemore, it is easy to notice that $(\partial K)_{\alpha\beta\gamma\delta}=1 $, hence it defines a cocycle, which represents the second Stiefel-Whitney class. In particular $w_2=0$ translates to
       \begin{equation*}
           [K]=\{ K_{\alpha\beta\gamma}\cdot (\partial \lambda)_{\alpha\beta\gamma}\hspace{1mm}|\hspace{1mm} \lambda_{\alpha\beta}\colon U_{\alpha\beta}\rightarrow \mathbb{Z}_2\}=1
       \end{equation*}
    It is clear that $[K]=1$ iff $K=\partial \lambda$. Defining $\tilde{g}'_{\alpha\beta}:=\lambda^{-1}_{\alpha\beta} \cdot g_{\alpha\beta} $, it is easy to show that $\tilde{g}'_{\alpha\beta}$ is a cocycle (i.e. satisfy the cocycle identity), hence it is possible to reconstruct a Spin-bundle from $(\mathcal{U},\{\tilde{g}'_{\alpha\beta}\})$, showing there are no obstructions for the existence of a spin structure.

    Conversely, if one assumes that a spin structure exists, then it is immediate to see $[K]=1$ because the lifted transition functions automatically satisfy the cocycle identity.

    We are only left with showing that $[K]$ is independent of the choice of trivialization and on the choice of the lift. We start by showing the independence on the choice of lift of $\{g_{\alpha\beta}\}$. Let $\kappa_{\alpha\beta}$ be some 1--cochain inducing $\tilde{g}"_{\alpha\beta}=\tilde{g}_{\alpha\beta}\cdot \kappa_{\alpha\beta}$. This defines a new lift $K"$ which is in the same equivalence class as $K$, as $K"=K\cdot \partial\kappa$.

    Now we can choose different local sections $\{g'_{\alpha\beta}\}$ for the original SO--bundle. We have previously seen that $g'_{\alpha\beta}=g_{\alpha}^{-1}\cdot g_{\alpha\beta}\cdot g_\beta$, which, after choosing a lift $\tilde{g}_\alpha$, gives $K'_{\alpha\beta\gamma}= \tilde{g}^{-1}_\beta\cdot  K_{\alpha\beta\gamma}\cdot\tilde{g}_\beta$. Now, since $K'_{\alpha\beta\gamma}\in\mathbb{Z}_2$, then $\tilde{g}_\beta$ and its inverse are both either $1$ or $-1$, so $K'_{\alpha\beta\gamma}=K_{\alpha\beta\gamma}$.
       
\end{proof}

As one can notice, so far we needed to fix a metric in order to define a spin strucutre. Now, we introduce an equivalent approach that does not rely on such assumption, and therefore is more suitable to work with theories where the metric is a dynamical field.

\begin{definition}
    Given a principal Spin$(d,1)$--bundle $P_s$, a spinbein map is a principal bundle morphism $\hat{e}:P_S\rightarrow LM$ satisfying verticality and equivariance, i.e. the following diagrams commute
    \begin{equation*}
\begin{tikzcd}
	{P_{s}} && LM && {P_{s}} & LM \\
	& M &&& {P_{SO}} & LM
	\arrow["{\tilde{e}}", from=1-1, to=1-3]
	\arrow["{\pi}", from=1-1, to=2-2]
	\arrow["{\pi}"', from=1-3, to=2-2]
	\arrow["{\tilde{e}}", from=1-5, to=1-6]
	\arrow["{\cdot \Lambda}", from=1-5, to=2-5]
	\arrow["{\cdot \Lambda}", from=1-6, to=2-6]
	\arrow["{\tilde{e}}", from=2-5, to=2-6]
\end{tikzcd}
    \end{equation*}
\end{definition}
As before, the spinbein defines a moving frame on sections $\hat{s}_\alpha:U_\alpha\rightarrow P_s$ as $\hat{e}(\hat{s}_\alpha(x))=(x,{}_\alpha( e_a))$, where ${}_\alpha e_a={}_\alpha e_a^\mu \partial_\mu$. On intersections the frames change by right action of an orthogonal transformation seen as the image under $l$ of a Spin transformation $S_{\alpha\beta}$ defining the transition functions, i.e.
    \begin{equation*}
        \hat{e}(\hat{s}_{\beta})=\hat{e}(\hat{s}_{\alpha})\cdot S_{\alpha\beta} \hspace{2mm} \Rightarrow \hspace{2mm} {}_{\beta}(e_a)={}_{\alpha}(e_b)l^{b}_a(S_{\alpha\beta}).
    \end{equation*}
\begin{remark}
    Also in this case, by dualizing the frames, one can induce unquely a metric on $M$ as $g_{\mu\nu}=e_\mu^a e_\nu^b \eta_{ab}$. Exactly as before, the image of $P_s$ under $\hat{e}$ turns out to be the orthogonal bundle $SO(M,g)$. Furthermore, lifting $l$ to a bundle map $\hat{l}\colon P_s\rightarrow P_{SO}$, it is clear that a trivialization on $P_s$ induces one on $P_{SO}$ via $\hat{l}$, and for each family of sections $\hat{s}_\alpha$ we obtain sections $s_\alpha:= \hat{l}\circ \hat{s}$. Equivalently, one obtains that the following diagram commutes
\begin{equation}\label{spin fram factorization}
	\begin{tikzcd}
		P_s \arrow[rdd, "\hat{p}", bend right] \arrow[rr, "\hat{e}"] \arrow[rd, "\hat{l}"] &                                  & LM \arrow[ldd, "\pi"', bend left] \\
		& P_{SO} \arrow[d, "p"] \arrow[ru, "e"] &                                   \\
		& M                                &                                  
	\end{tikzcd}
\end{equation}
Notice that also a vielbein map $e$ equivalent to the one introduced in the previous chapter is introduced. 
\end{remark}    
The reason for the last statement is clear when one defines the associated bundle
    \begin{equation*}
        \hat{\mathcal{V}}:=P_s\times_{\hat{\rho}} V,
    \end{equation*}
where $\hat{\rho}$ is the vector (i.e. spin 1) representation of Spin$(d,1)$ on V. Notice however how every integer spin representation $\hat{\lambda}$ of Spin$(d,1)$ is the same as a representation of SO$(d,1)$, as it factors through the double cover $\hat{\lambda}=\lambda\circ l$. In particular, this tells us that $\hat{\mathcal{V}}\simeq \mathcal{V}$ and that a spin coframe 
    \begin{equation*}
        e\colon TM \overset{\sim}{\longrightarrow} \hat{\mathcal{V}}
    \end{equation*}
produces the same dynamics as the vielbein field. The advantage of using spin bundles is that of being able to define associated vector bundles with respect to half-integer spin representations, i.e. spinor bundles.

\begin{theorem}\cite{FatiNoris22}
    A spinbein map $\hat{e}$ on $M$ exists if and only if a spin structure exists on $(M,g)$ for some metric $g$.
\end{theorem}
\begin{proof}
    Given a spinbein map $\hat{e}\colon P_s \rightarrow LM$, it induces a spin structure just by restricting the target to the image of $\hat{e}$, i.e.
        \begin{equation*}
            \Sigma\colon P_s \overset{\hat{e}}{\longrightarrow} \hat{e}(P_s)=SO(M,g),
        \end{equation*}
    where in this case $g$ is the metric induced by the coframe defined by $\hat{e}$. Conversely, if $\Sigma:P_s\rightarrow SO(M,g)$ is a spin structure, one can induce a spinbein map $\hat{e}:=\hat{\iota}\circ \Sigma$, where $\hat{\iota}:P_{SO}\rightarrow LM$ is the inclusion in the frame bundle.    
\end{proof}

Having proved this, it is clear that using spin coframes is allowed exactly when spin structures exist, and viceversa, hence we can regard it as an equivalent description. 

Finally, we have all the ingredients to define spinor bundles.
\begin{definition}
    Let $V_{\mathbb{C}}$ be the complexification of the $D$-dimensional real vector space $V$. By the discussion in the previous chapter, we know that, depending on the parity of $D$, there exist faithful representations of the Clifford algebra $\mathcal{C}(V_{\mathbb{C}})=\mathcal{C}(V)_{\mathbb{C}}$. In particular, we are interested in the gamma representation $\Gamma$ of proposition \ref{prop: cmplx clif lorentz gamma}, which allows to define the Dirac spinor bundle as     
    \begin{equation*}
        \mathbb{S}_D:=P_S \times_\Gamma \mathbb{C}^{2^{\frac{D}{2}}}
    \end{equation*}
    Sections of $\mathbb{S}_D$ are called Dirac spinor fields. Furthermore, when the dimension allows it, one can also define the subbundle of Majorana spinors as 
        \begin{equation*}
            \mathbb{S}_{M}:=\bigcup_{x\in M} \{(x,\psi)\in \mathbb{S}_{D,x}\hspace{1mm}\vert\hspace{1mm} C \Gamma_0 \psi^*=\psi  \}.
        \end{equation*}
\end{definition}

\section{Tools and identities}\label{sec: identities}

    \subsection{Basic Identities on gamma matrices}
        Let $a=0,\cdots,d$. Setting $\Gamma_{a_1\cdots a_n}:= \Gamma_{[a_1} \Gamma_{a_2}\cdots \Gamma_{a_n]}  $, we present a list of well known identities\footnote{which the reader can easily check by permuting the gamma matrices using their defining equations} adjusted to the mostly plus signature:

    \begin{align}
        &\label{id:gamma2} \Gamma^a\Gamma_a=-D;\\
        &\label{id:gamma3}  \Gamma^a\Gamma^b\Gamma_a=(D-2)\Gamma^b;\\
        &\label{id:gamma4}  \Gamma^a\Gamma^b\Gamma^c\Gamma_a=(4-D)\Gamma^b\Gamma^c + 4 \eta^{bc}\mathbb{1};\\
        &\label{id:gamma5} \Gamma^a\Gamma^b\Gamma^c\Gamma^d\Gamma_a=(D-6)\Gamma^b\Gamma^c\Gamma^d - 4 \eta^{cd}\Gamma^b - 4 \eta^{bc}\Gamma^d + 4 \eta^{bd}\Gamma^c ;\\
        &\label{id:easy shuffle ganmas} \Gamma^{a_1\cdots a_r}=\frac{1}{2}\left( \Gamma^{a_1}\Gamma^{a_2\cdots a_r} - (-1)^r \Gamma^{a_2\cdots a_r} \Gamma^{a_1} \right);\\
        &\label{id:anti gamma a r a} \Gamma^a \Gamma^{a_1 \cdots a_r}\Gamma_a =  (-1)^{r+1}(D-2r)\Gamma^{a_1 \cdots a_r};\\
        &\label{id: contraction s gamma's} \Gamma^{a_1\cdots a_r b_1\cdots b_s}\Gamma_{b_1 \cdots b_s}=(-1)^r \frac{(D-r)!}{(D-r-s)!} \Gamma^{a_1\cdots a_r }  
    \end{align}
    
In $D=4$, where we denote gamma matrices with $\{\gamma_a\}$, having set $\gamma^5:=i\gamma^0\gamma^1\gamma^2\gamma^3$, the following identities hold:
    \begin{align}
        &\label{id:d=4 gamma5} \gamma^a\gamma^b\gamma^c\gamma^d\gamma_a=2\gamma^d\gamma^c\gamma^b;\\
        &\label{id:d=4 gamma3} \gamma^a\gamma^b\gamma^c=-\eta^{ab}\gamma^c-\eta^{bc}\gamma^a+\eta^{ac}\gamma^b + i \epsilon^{dabc}\gamma_d\gamma^5;\\
        &\label{id:d=4 gamma5,2}\gamma^5\gamma^{cd}=-\frac{i}{2}\epsilon^{abcd}\gamma_{ab};\\
        &\label{id: gamma5gammac} \gamma^a\gamma^5=i\epsilon^{abcd}\gamma_{bcd}.
    \end{align}
    Considering $\{v_a\}$ basis for $V$, we set $\Gamma:=\Gamma^a v_a$\footnote{From now on we will omit the $\wedge$ symbol and automatically assume that for all $B\in V$, $B^N=B\wedge\cdots\wedge B\in \wedge^N V$.} and define the bracket $[\cdot,\cdot]$ to encompass the action of $\mathfrak{spin}(d,1)\simeq \wedge^2 V$ on $\wedge^j V$, i.e extend by linearity and graded Leibniz (on the first and second entries) the following
        \begin{align*} 
            [v_a,\cdot]\colon &\wedge^k V \longrightarrow  \wedge^{k-1}V\\
            &\alpha=\frac{1}{k!}\alpha^{a_1\cdots a_k}v_{a_1}\cdots v_{a_k} \longmapsto\frac{1}{(k-1)!}\eta_{a a_1} \alpha^{a_1\cdots a_k} v_{a_2}\cdots v_{a_k}
        \end{align*}
    we obtain 
        \begin{align}
            & \label{id:v_a,gamma^N}[v_a,\Gamma^N]=N[v_a,\Gamma]\Gamma^{N-1} + N(N-1)v_a\Gamma^{N-2} ,\quad N\geq 2;\\
            &\label{id:v_a,gamma^N2}\hspace{11.5mm}=(-1)^{N-1}(N\Gamma^{N-1}\Gamma_a + N(N-1)\Gamma^{N-2}v_a).
        \end{align}
        
    Now we are interested in the cases when the expression containing spinors is real (whether it is because it contains Majorana-type spinors or because we are dealing with real quantities defined via Dirac spinors). In particular, in most of the relevant computations, denoting  complex conjugation by $(\cdot)^*$, one considers $iA-iA^*$, where $A$ is any expression containing spinors. Here we list some recurring expressions:
        \begin{align}
            & \label{id:v_a,gamma,gamma^N}[v_a,\Gamma] \Gamma^N  - (-1)^N \Gamma^N [v_a,\Gamma]  = -2Nv_a\Gamma^{N-1};\\
            & \label{id:theta,gamma,gamma^2} [\Gamma,\Theta]\Gamma^2 - \Gamma^2[\Gamma,\Theta] =4N\Gamma\Theta \quad \forall \\
            &\label{id:gamma^3 godplz}\bar{\chi}\gamma^3[\alpha,\psi]=3\bar{\chi}\gamma\psi + (-1)^{|\alpha|}\frac{1}{2}\bar{\chi}[\alpha,\gamma^3]_v\psi,
        \end{align}
        for all $ \alpha \in \wedge^2 V, \theta\in\wedge^{D-3} V$, $\Theta\in \wedge^N V$ and even Majorana spinors $\chi$ and $\psi$, having defined
            \begin{equation*}
            [\alpha,\psi]:=\frac{1}{4}[\gamma,[\gamma,\alpha]]_V\psi=-\frac{1}{4}\alpha^{ab}\gamma_{ab}\psi,
            \end{equation*}
        having considered the split $[\alpha,\gamma]=[\alpha,\gamma]_{\mathcal{C}}+[\alpha,\gamma]_V=0$, since an element in $\wedge^2 V\simeq \mathfrak{spin}(d,1)$ acts both via the Gamma representation and on $V$ via the fundamental representation.    
    \subsection{Identities on Majorana spinors}
    \subsubsection{Majorana flip relations}
            Let $D=2k,2k+1$. Given any two Majorana spinors $\psi$ and $\chi$ (for which, we recall, $\epsilon=1,\eta=-1$)  of arbitrary parity, denoting $\Gamma=\Gamma^a v_a\in \mathrm{End}(\mathbb{C}^{2^k})\otimes V$, we have the following 
                \begin{align}
                    &  \label{flip:0}\bar{\chi}\psi= -(-1)^{|\chi||\psi|}\bar{\psi}\chi;\\
                    &  \label{flip:1}\bar{\chi}\Gamma\psi=(-1)^{|\psi|+|\chi|+|\psi||\chi|}\bar{\psi}\Gamma\chi; \\
                    & \label{flip:2} \bar{\chi}\Gamma^2\psi=(-1)^{|\psi||\chi|}\bar{\psi}\Gamma^2\chi  ;\\
                    &   \label{flip:3}\bar{\chi}\Gamma^3\psi=-(-1)^{|\psi|+|\chi|+|\psi||\chi|}\bar{\psi}\Gamma^3\chi; 
                \end{align}
    
            In general, one finds 
                \begin{equation}
                    \label{flip:N}\bar{\chi}\Gamma^N\psi=-t_N(-1)^{N(|\psi|+|\chi|)+|\psi||\chi|}\bar{\psi}\Gamma^N\chi,
                \end{equation}
            where $t_N$ is defined from $(C\Gamma^N)^t=-t_NC \Gamma^N$ and is such that $t_{N+4}=t_N$.\footnote{A closer inspection reveals $t_0=t_3=-\epsilon \eta = 1$, $t_1=t_2=-\epsilon$,} The first 4 parameters read
                \begin{equation*}
                    t_0=1,\qquad t_1=-1, \qquad t_2=-1, \qquad t_3=1,
                \end{equation*}
            while the general formula is
                \begin{equation}
                    t_N=(-1)^{\floor*{\frac{N+1}{2}}}.
                \end{equation}
            \subsubsection{Fierz identities}
            As stated in \eqref{Clifford basis}, one can find a basis of the Clifford algebra using products of elements of the basis of $V$. In the context of the gamma representation, the Clifford product is mapped into matrix multiplication, hence a basis is given in terms of products of gamma matrices as  $\{\Gamma^{[A]}\}:=\{ \mathbb{1}, \Gamma^a, \Gamma^{ab} ,\cdots, \Gamma^{a_0\cdots a_d}  \}$, where $[A]$ represents the number of factors in the basis element, also known as the rank of the basis element. We define $\{\Gamma_{[A]}\}:==\{ \mathbb{1}, \Gamma_a, \Gamma_{ba} ,\cdots, \Gamma_{a_d\cdots a_0}  \}$ with lower indices in the opposite order, as it helps with signs arising in the computations.

            Starting by the even dimensional case where $D=2k$, we aim at using the generators $\{\Gamma^{[A]}\} $ to obtain any matrix on $\mathbb{C}(2^k) $. Indeed, on $\mathbb{C}(2^k) $, one has the obvious pairing introduced by the trace operator, i.e. $\forall M,N\in\mathbb{C}(2^k)$, $(M,N):=\Tr(M N^\dag)$. 

            It can be shown that, for even dimensions $D=2k$, one has the following property
            \begin{equation}
                \Tr(\Gamma^{[A]}\Gamma_{[B]})=(-1)^{[A]}2^k\delta^{[A]}_{[B]},
            \end{equation}  
            where for a generic index set $[A]=a_1\cdot a_r$, $\delta^{[A]}_{[B]}:=\delta^{a_1\cdots a_r}_{b_1\cdots b_r}:=\delta^{[a_1]}_{b_1}\cdots\delta^{a_r]}_{b_r}  $.  The above relation allows to expand any matrix $M\in\mathbb{C}(2^k)$ as a linear combination of products of gamma matrices, i.e.
                \begin{equation*}
                    M=\sum_A^{} m_{[A]}\Gamma^{[A]} \qquad \text{with} \qquad m_{[A]}=\frac{(-1)^{[A]} }{2^m}\Tr(M\Gamma_{[A]}).
                \end{equation*}
            Denoting by $\alpha=1,\cdots,2^k$ the spinor indices, following \cite{Freedman:2012zz}, one can consider $\delta^{\beta}_{\alpha}\delta^\delta_\gamma$ as a matrix with entries labelled by indices $\beta$ and $\gamma$, while $\alpha$ and $\delta$ are just dummy inert indices. Applying the above formula we obtain   
                \begin{align*}
                    \delta^{\beta}_{\alpha}\delta^\delta_\gamma=\sum_A (m_{[A]})^\delta_\alpha(\Gamma^{[A]})^\beta_\gamma, \qquad(m_{[A]})^\delta_\alpha = \frac{(-1)^{[A]} }{2^k} \delta^{\beta}_{\alpha}\delta^\delta_\gamma (\Gamma_{[A]})^\gamma_\beta=\frac{(-1)^{[A]} }{2^k}(\Gamma_{[A]})^\delta_\alpha,
                \end{align*}
            obtaining
                \begin{equation*}
                    \delta^{\beta}_{\alpha}\delta^\delta_\gamma=\sum_A \frac{(-1)^{[A]} }{2^k}(\Gamma_{[A]})^\delta_\alpha (\Gamma^{[A]})^\beta_\gamma
                \end{equation*}
            We are interested in the decomposition of $\gamma^a \gamma_a$ in even dimensions. We obtain 
                \begin{align*}                            (\Gamma^a)^\rho_\alpha(\Gamma_a)^\delta_\sigma&=(\Gamma^a)^\rho_\beta(\Gamma_a)^\gamma_\sigma\delta^{\beta}_{\alpha}\delta^\delta_\gamma\\
                 &=(\Gamma^a)^\rho_\beta(\Gamma_a)^\gamma_\sigma\sum_A \frac{(-1)^{[A]} }{2^k}(\Gamma_{[A]})^\delta_\alpha (\Gamma^{[A]})^\beta_\gamma\\
                    &=\sum_A \frac{(-1)^{[A]} }{2^k}(\Gamma_{[A]})^\delta_\alpha (\Gamma^a\Gamma^{[A]}\Gamma_a)^\rho_\sigma.
                \end{align*}
            From  \eqref{id:anti gamma a r a}, we see $\Gamma^a \Gamma^{[A]}\Gamma_a=(-1)^{[A]+1}(D-2[A])\Gamma^{[A]} $, hence obtaining 
            \begin{equation}
                (\Gamma^a)^\rho_\alpha(\Gamma_a)^\delta_\sigma=\sum_A \frac{(2[A]-D)}{2^k}(\Gamma^{[A]})^\rho_\sigma(\Gamma_{[A]})^\delta_\alpha 
            \end{equation}
            
            Now we consider the case $D=4$. We can use the charge conjugation matrix to lower the indices of the gamma matrices\footnote{It is also useful to adopt this formalism when dealing with scalar quantities defined in terms of spinors. For example, we have $\bar{\chi}_\alpha=\chi^\beta C_{\alpha\beta}$ and $C_{\beta\alpha}:= \delta_{\epsilon \alpha} C^\epsilon_\beta$. 
            } and obtain $(\gamma^a)^\cdot_{\alpha \beta}:=C_{\alpha \delta}(\gamma^a)^\delta_\beta.$ Furthermore, we symmetrize the part in $(\beta \rho\delta)$ obtaining
                \begin{align*}
                    (\gamma^a)^\cdot_{\alpha(\beta}(\gamma^a)^\cdot_{\rho\delta)}&=\frac{1}{4}\left[- 4 C_{\alpha(\delta}C_{\rho\beta)} -2(\gamma^a)^\cdot_{\alpha(\delta}(\gamma^a)^\cdot_{\rho\beta)} - 0 +2 (\gamma^{abc})^\cdot_{\alpha(\delta}(\gamma_{abc})^\cdot_{\rho\beta)} + 4  (\gamma^5)^\cdot_{\alpha(\delta}(\gamma^5)^\cdot_{\rho\beta)} \right]\\
                    &=-\frac{1}{2}(\gamma^a)^\cdot_{\alpha(\delta}(\gamma^a)^\cdot_{\rho\beta)}=0,
                \end{align*}
            Having used that $C_{(\alpha\beta)}=0$, $(\gamma_{abc})^\cdot_{(\rho\beta)}=0$ and that $(\gamma^5)^\cdot_{(\rho\beta)}=0$, as a consequence of the fact that 
                \begin{equation}
                    (\gamma^{a_1\cdots a_r})^\cdot_{\alpha\beta}=-t_r(\gamma^{a_1\cdots a_r})^\cdot_{\beta\alpha}.
                \end{equation}
            Then one finds
                \begin{equation}\label{Fierz:0}
                    (\gamma^a)^\cdot_{\alpha (\beta}(\gamma_a)^\cdot_{\rho \delta)} =0   .
                \end{equation}
    Contracting with 4 Majorana spinors $\lambda_i$'s ($i=1, \cdots, 4$) of arbitary parity, we obtain 
                \begin{align}
                    & \label{Fierz:1}\bar{\lambda}_1 \gamma^3 \lambda_2 \bar{\lambda}_3 \gamma \lambda_4=(-1)^{|\lambda_2||\lambda_3|}\bar{\lambda}_1\gamma\lambda_3 \bar{\lambda}_2 \gamma^3 \lambda_4 + (-1)^{|\lambda_4|(|\lambda_2|+|\lambda_3|+1)+|\lambda_3|} \bar{\lambda}_1\gamma\lambda_4 \bar{\lambda}_2 \gamma^3 \lambda_3;\\
                    & \label{Fierz:2} \bar{\lambda}_1 \gamma^3 \lambda_2 \bar{\lambda}_3 \gamma \lambda_4= - (-1)^{|\lambda_2||\lambda_3|}\bar{\lambda}_1\gamma^3\lambda_3 \bar{\lambda}_2 \gamma \lambda_4 - (-1)^{|\lambda_4|(|\lambda_2|+|\lambda_3|+1)+|\lambda_3|} \bar{\lambda}_1\gamma^3\lambda_4 \bar{\lambda}_2 \gamma \lambda_3 .
                \end{align}
                
\subsection{Lemmata and other facts about coframes}
 We start by defining the space
        \begin{equation*}
            \Omega^{(k,l)}:=\Omega^k(M,\wedge^l \mathcal{V}),
        \end{equation*}
and the maps
    \begin{align*}
        W_e^{(i,j)}\colon & \Omega^{(i,j)}\rightarrow \Omega^{(i+1,j+1)}\\
        &\alpha\mapsto e\wedge \alpha,
    \end{align*}
where $\mathcal{V}$ is identified with $\hat{\mathcal{V}}$ and $e$ is a spin coframe. From now on, we also omit the wedge symbol and assume it between differential forms. The properties of these maps (including the case when the coframes are restricted to the boundary and corners) are found in \cite{Canepa:2024rib}.

We then obtain the following results in $D=4$.
    \begin{lemma}\label{lem: injectivity egamma^3}
        The map 
        \begin{align*}
            \Theta^{(1,0)}\colon  \Omega^{(1,0)}(\Pi \mathbb{S}_D)& \longrightarrow \Omega^{(2,4)}(\Pi \mathbb{S}_D)\\
            \psi& \longmapsto \frac{1}{3!}e\gamma^3 \psi
        \end{align*}
        is injective.
    \end{lemma}
    \begin{proof}Using $v_a v_b v_c v_d = \epsilon_{abcd} \mathrm{Vol}_V$
        \begin{align*}
            \frac{1}{3!}e\gamma^3\psi&=\frac{1}{3!}e^a\gamma^{bcd}\psi v_av_bv_cv_d\\
            &=\frac{1}{3!}\epsilon_{abcd} e^a\gamma^{bcd} \psi \mathrm{Vol}_V\\
            &\overset{\eqref{id: gamma5gammac}}{=}\frac{1}{3!}i\gamma^5 e^a\gamma_a \psi\mathrm{Vol}_V=0 \qquad \Leftrightarrow \qquad [e,\gamma]\psi=0.
        \end{align*}
    Now $[e,\gamma]\psi=0$ if and only if $\gamma_{[\mu}\psi_{\nu]}=0$, which is uniquely solved by $\psi=0,$ hence proving $ \Theta^{(1,0)}$ is injective.
    \end{proof}
        \begin{lemma}\label{lem: iso (1,0) (3,4)}
        The map 
        \begin{align*}
            \Theta_\gamma^{(1,0)}\colon  \Omega^{(1,0)}(\Pi \mathbb{S}_D)& \longrightarrow \Omega^{(3,4)}(\Pi \mathbb{S}_D)\\
            \psi& \longmapsto \frac{1}{3!}e\gamma^3 \underline{\gamma}\psi
        \end{align*}
        is an isomorphism, where $\underline{\gamma}:= [e,\gamma]=\gamma_\mu dx^\mu$
    \end{lemma}
    \begin{proof}
        First of all, from the previous proof we know $e \gamma^3 = i\gamma^5 [e,\gamma]=i\gamma^5 \ugam \mathrm{Vol}_V$. Then
        \begin{align*}
            e\gamma^3 \ugam \psi = i \gamma^5 \ugam^2 \psi \mathrm{Vol}_V=0 \quad \Leftrightarrow\quad \ugam^2 \psi =0 \quad \Leftrightarrow\quad  \gamma_{[\mu} \gamma_\nu \psi_{\rho]}=0.
        \end{align*} 
        The latter is a system of 4 equations whose solution (due to invertibility of the gamma matrices) is uniquely given by $\psi_\rho=0$. This shows that $\Theta_\gamma^{(1,0)}$ is injective, but since $\dim \Omega^{(1,0)}=\dim \Omega^{(3,4)}$ and $\Theta_\gamma^{(1,0)}$ is linear, by the rank theorem $\dim\mathrm{Im}(\Theta_\gamma^{(1,0)})=\dim\Omega^{(3,4)} $ hence it is also surjective. 
    \end{proof}
    \begin{remark}
        By the same reasoning (or just by taking the Dirac conjugate of the above expression), one finds that also the map
            \begin{equation*}
                 \psi \longmapsto \frac{1}{3!}e\ugam\gamma^3 \psi
            \end{equation*}
        is an isomorphism.    
    \end{remark}
    \begin{lemma}\label{lem: splitting (3,1)}
        For all $\theta \in \Omega^{(3,1)}(\Pi \mathbb{S}_M) $ there exist unique $\alpha\in \Omega^{(1,0)}(\Pi \mathbb{S}_M) $ and $\beta \in \Omega^{(3,1)}(\Pi \mathbb{S}_M) $ such that
            \begin{equation}
                \theta = i e \underline{\gamma} \alpha + \beta \qquad \mathrm{and}\qquad \gamma^3 \beta = 0.
            \end{equation}
    \end{lemma}
    \begin{proof}
        We start by considering the map $(e\ugam)_{(1,0)}\colon\Omega^{(1,0)}\rightarrow \Omega^{(3,1)} \colon \alpha \mapsto e\ugam\alpha $. We see that $e\ugam\alpha=0$ implies $\ugam\alpha+0$ due to injectivity of $W_e^{(1,0)}$\cite{Canepa:2024rib}, while
            \begin{equation*}
                \ugam\alpha=0 \quad \Leftrightarrow \quad \gamma_{[\mu}\alpha_{\nu]}=0 \quad \Leftrightarrow \quad \alpha_\nu=0,
            \end{equation*}
        hence implying that $(e\ugam)_{(1,0)}$ is injective. 

        Now, defining $(\gamma^3)_{(3,1)}\colon\Omega^{(3,1)}\rightarrow\Omega^{(3,4)}\colon\beta\mapsto\gamma^3\beta $, we notice that $\gamma^3\beta=\mathrm{Vol}_V [\gamma,\beta]_V$, hence $\ker((\gamma^3)_{(3,1)})=\{ \beta\in\Omega^{(3,1)}\hspace{1mm}|\hspace{1mm}[\gamma,\beta]=0 \}$. We have
            \begin{equation*}
                [\gamma,\beta]=\gamma_a\beta^a_{\mu\nu\rho}=0,
            \end{equation*}
        which is a system of 4 independent equations, implying $\dim(\ker((\gamma^3)_{(3,1)}))=\dim(\Omega^{(3,1)})-4=12$. Now, since   $(e\ugam)_{(1,0)}$ is injective, it is immediate to see that $$\dim(\Omega^{(3,1)})=16=\dim(\mathrm{Im}((e\ugam)_{(1,0)}))+\dim(\ker((\gamma^3)_{(3,1)}))=\dim(\Omega^{(1,0)})+\dim(\ker((\gamma^3)_{(3,1)})).$$ The claim is then proved once we show that $\mathrm{Im}((e\ugam)_{(1,0)})\cap\ker((\gamma^3)_{(3,1)})=\{0\} $. This is immediate since, by lemma \ref{lem: iso (1,0) (3,4)}, for all $\alpha\in\Omega^{(1,0)}$, 
            \begin{equation*}
                \gamma^3e\ugam\alpha=0\quad\Leftrightarrow\quad \alpha=0.
            \end{equation*}
    \end{proof}
    \begin{lemma}\label{lem: splitting (2,1)}
        Let $n\in\mathbb{N}$ and $\ugam\gamma^n_{(i,j)}$ be the map
            \begin{equation*}
                \ugam\gamma^n_{(i,j)}\colon\Omega^{(i,j)}\rightarrow \Omega^{(i,j+n)}\colon\beta\mapsto\ugam \gamma^n \beta.
            \end{equation*}
        Then, for all $\theta \in \Omega^{(2,1)}(\Pi \mathbb{S}_M)$ there exist unique $\alpha\in\Omega^{(1,0)}(\Pi \mathbb{S}_M)$ and $\beta \in \ker{\ugam\gamma^3_{(2,1)}}$ such that 
            \begin{equation*}
                \theta= e \alpha + \beta. 
            \end{equation*}
    \end{lemma}
    \begin{proof}
        We see $\dim\mathrm{Im}(W_1^{(1,0)})= \dim\Omega^{(1,0)}=4$ as $W_1^{(1,0)}$ is injective. On the other hand, from \ref{lem: iso (1,0) (3,4)}, we see $\Omega^{(3,4)}=e\ugam\gamma^3\Omega^{(1,0)}$, implying in particular that $\ugam\gamma^3_{(2,1)}$ is surjective, hence $\dim\ker(\ugam\gamma^3_{(2,1)})=\dim\Omega^{(2,1)} - \dim \Omega^{(3,4)}=20$. Now, since $\dim\Omega^{(2,1)}=\dim\mathrm{Im}(W_1^{(1,0)})+\dim\ker(\ugam\gamma^3_{(2,1)})$, we just need to prove that $\mathrm{Im}(W_1^{(1,0)})\cap \ker(\ugam\gamma^3_{(2,1)})=\{0\}$. Choosing $\alpha\in \Omega^{(1,0)}$, we see 
            \begin{equation*}
                e \alpha \in \ker(\ugam\gamma^3_{(2,1)}) \quad \Leftrightarrow \quad e\ugam\gamma^3 \alpha=0 \quad \Leftrightarrow \quad \alpha=0.
            \end{equation*}
        For uniqueness, assume there exist $\alpha_1,\alpha_2\in\Omega^{(1,0)}$ and $\beta_1,\beta_2\in\ker(\ugam\gamma^3_{(2,1)})$ such that $\theta=e\alpha_1 +\beta_1=e\alpha_2 +\beta_2$, then 
            \begin{equation*}
                e(\alpha_1-\alpha_2)=\beta_2-\beta_1\in \ker(\ugam\gamma^3_{(2,1)}) ,
            \end{equation*}
        which implies $\alpha_1 - \alpha_2=0$, and $\beta_2-\beta_1=0$.     
    \end{proof}

    \begin{lemma}\label{lem: Fierz}
        For all $\lambda,\psi,\chi\in \mathbb{S}_M$  such that $|\chi|=0$ and $|\psi|=1$, the following identities hold
            \begin{equation*}
                \bar{\lambda}\gamma^3 \chi \bar{\chi}\gamma\psi =0, \qquad  \bar{\chi}\gamma\chi \bar{\lambda}\gamma^3 \psi=0 \qquad \mathrm{and} \qquad \bar{\lambda}\gamma\chi \bar{\chi}\gamma^3\psi=0.
            \end{equation*}
    \end{lemma}
    \begin{proof}
        The proof of the above identity rely on subsequent applications of Fierz identities \eqref{Fierz:1} and \eqref{Fierz:2} and Majorana flip relations. In particular $\bar{\lambda}\gamma^3 \chi \bar{\chi}\gamma\psi \overset{\eqref{Fierz:1}}{=} \bar{\lambda}\gamma \chi \bar{\chi}\gamma^3\psi + (-1)^{|\psi|}\bar{\lambda}\gamma\psi \bar{\chi}\gamma^3\chi =  \bar{\lambda}\gamma \chi \bar{\chi}\gamma^3\psi$, having used \eqref{flip:3}. At the same time
        \begin{align*}
             \bar{\lambda}\gamma \chi \bar{\chi}\gamma^3\psi & \overset{\eqref{flip:3}}{=} (-1)^{|\lambda|(|\psi|+1)}\bar{\chi}\gamma^3 \psi\bar{\lambda}\gamma\chi\\
             &\overset{\eqref{Fierz:1}}{=}(-1)^{|\lambda|(|\psi|+1)}\left( (-1)^{|\lambda||\psi|}\bar{\chi}\gamma\psi\bar{\psi}\gamma^3\chi + (-1)^{|\lambda|}\bar{\chi}\gamma\chi \bar{\lambda}\gamma^3\psi \right)\\
             &\overset{\eqref{flip:1}\eqref{flip:3}}{=}-(-1)^{|\psi|}\bar{\lambda}\gamma \chi \bar{\chi}\gamma^3\psi + (-1)^{|\lambda||\psi|}\bar{\chi}\gamma\chi \bar{\lambda}\gamma^3\psi,
        \end{align*}
        hence showing that when $|\psi|=1$, $\bar{\chi}\gamma\chi \bar{\lambda}\gamma^3\psi=0 $. Now at the same time we have
            \begin{equation*}
                \bar{\lambda}\gamma^3 \chi \bar{\chi}\gamma\psi \overset{\eqref{Fierz:2}}{=} - \bar{\lambda}\gamma^3\chi - (-1)^{|\psi|}\bar{\lambda}\gamma^3\psi \bar{\chi}\gamma\chi,
            \end{equation*}
        hence $\bar{\lambda}\gamma^3 \chi \bar{\chi}\gamma\psi  = - \frac{1}{2}(-1)^{|\psi|}\bar{\lambda}\gamma^3\psi \bar{\chi}\gamma\chi=0$. Lastly, we saw that $\bar{\lambda}\gamma \chi \bar{\chi}\gamma^3\psi=\bar{\lambda}\gamma^3 \chi \bar{\chi}\gamma\psi =0.$    
    \end{proof}

\appendix

\section{Proofs of section \ref{sec: identities}}

        We now list the proofs of equations in \ref{sec: identities}
        
        \begin{itemize}            
            \item \eqref{id:v_a,gamma^N} and \eqref{id:v_a,gamma^N2}. We prove this by induction, first showing it holds for $N=2,3$ and then proving the inductive step, having set $\Gamma_a:=\eta_{ab}\Gamma^b=[v_a,\Gamma]$.
            \begin{align*}
                [v_a,\Gamma^2]&=\Gamma_a \Gamma - \Gamma \Gamma_a= \Gamma_a \Gamma - \Gamma^b v_b\Gamma^c \eta_{ac}= \Gamma_a \Gamma + \Gamma^c \eta_{ac}\Gamma^b v_b + 2 \eta^{bc} \eta_{ac} v_b = 2 \Gamma_a \Gamma + 2 v_a\\
                &=2\Gamma^c \Gamma^b \eta_{ac}v_b + 2v_a = - 2 \Gamma \Gamma_a - 4 \eta^{cb}\eta_{ac}v_b + 2v_a= - 2 \Gamma \Gamma_a - 2 v_a,
            \end{align*}
            \begin{align*}
                [v_a,\Gamma^3]&=[v_a,\Gamma^2]\Gamma + \Gamma^2 \Gamma_a = - 2 \Gamma^c \Gamma^d \Gamma^b \eta_{ad}v_c v_b - 2 v_a \Gamma + \Gamma^2 \Gamma_a\\
                &= 2 \Gamma^c \Gamma^b \Gamma^d \eta_{ad}v_c v_b + 4 \eta^{db}\eta_{ad}\Gamma^c v_c v_b + 2\Gamma v_a + \Gamma^2 \Gamma_a = 3 \Gamma^2 \Gamma_a + 6\Gamma v_a\\
                &=\Gamma_a \Gamma^2 - \Gamma [v_a,\Gamma^2]=\Gamma_a \Gamma^2 - 2\Gamma \Gamma_a \Gamma - 2 \Gamma v_a\\
                &=\Gamma_a \Gamma^2 + 2 \Gamma^b \Gamma^c \Gamma^d \eta_{ad} v_b v_c + 4 \eta^{cd}\eta_{ad}\Gamma^b v_b v_c + 2 v_a \Gamma\\
                &=3 \Gamma_a \Gamma^2 + 6 v_a \Gamma.
            \end{align*}
            Now assume \eqref{id:v_a,gamma^N} and \eqref{id:v_a,gamma^N2} hold for $N-1$, then
            \begin{align*}
                [v_a,\Gamma^N]&=\Gamma_a \Gamma^{N-1} - \Gamma [v_a,\Gamma^{N-1}]= \Gamma_a \Gamma^{N-1} - (N-1) \Gamma \Gamma_a \Gamma^{N-2} - (N-1)(N-2) \Gamma v_a \Gamma^{N-3}\\
                &= N \Gamma_a \Gamma^{N-1} + 2(N-1) v_a \Gamma^{N-2} + (N-1)(N-2)v_a \Gamma \Gamma^{N-2}\\
                &= N \Gamma_a \Gamma^{N-1} + N(N-1) v_a \Gamma^{N-2}\\
                &=[v_a, \Gamma^{N-1}]\Gamma +(-1)^{N-1}\Gamma^{N-1}\Gamma_a\\
                &= (-1)^{N-2}[(N-1)\Gamma^{N-2}\Gamma_a (N-1)(N-2)\Gamma^{N-3}v_a]\Gamma + (-1)^{N-1}\Gamma^{N-1}\Gamma_a\\
                &=(-1)^{N-1}(N\Gamma^{N-1}\Gamma_a + N(N-1)\Gamma^{N-2}v_a );
            \end{align*}
            \item \eqref{id:v_a,gamma,gamma^N} follows immediately by subtracting \eqref{id:v_a,gamma^N} from \eqref{id:v_a,gamma^N2} applied to $\Gamma^{N+1}$:
            \item\eqref{id:theta,gamma,gamma^2}. Consider $\Theta=\frac{1}{N!}\Theta^{a_1\cdots a_n} v_{a_1}\cdots v_{a_N} $, then 
                \begin{align*}
                    [\Gamma,\Theta]\Gamma^2&=\frac{(-1)^{|\Theta|-N}}{(N-1)!}\Theta^{a_1a_2\cdots a_N}v_{a_2}\cdots v_{a_N}\eta_{a_1a}\Gamma^a\Gamma^b\Gamma^c v_b v_c\\
                    &=-\frac{(-1)^{|\Theta|-N}}{(N-1)!}\Theta^{a_1a_2\cdots a_N}v_c v_bv_{a_2}\cdots v_{a_N}\eta_{a_1a}(-4\eta^{ab}\Gamma^c + \Gamma^b\Gamma^c\Gamma^a)\\
                    &=\Gamma^2[\Gamma,\Theta] + 4N \Gamma\Theta;
                \end{align*}
            \item \eqref{id:gamma^3 godplz} Consider $\alpha\in\wedge^2 V$ with parity $|\alpha|$, then for any Dirac spinor (of any parity) $\chi$ and $\psi$ we have $\bar{\chi}\gamma^3[\alpha,\psi]=\frac{1}{4}\bar{\chi}\gamma^3\gamma^a\gamma^b [v_a,[v_b,\alpha]]\psi$, so 
                \begin{align*}
                    \gamma^3\gamma^a\gamma^b [v_a,[v_b,\alpha]]&=-[v_a,\gamma^3[v_b,\alpha]]\gamma^a \gamma^b + (3\gamma^2 \gamma_a + 6 \gamma v_a)[v_b,\alpha]\gamma^a \gamma^b\\
                    &=-[v_a,(3\gamma^2\gamma_b + 6\gamma v_b)\gamma^a\gamma^b \alpha] - 6\gamma^2[\gamma,\alpha]_V\\
                    &= - 6\gamma^2[\gamma,\alpha]_V + [v_a, 12 \gamma v_b\eta^{ab}\alpha]\\
                    &=- 6\gamma^2[\gamma,\alpha]_V + 36\gamma \alpha + (-1)^{|\alpha|}12\gamma v_b \eta^{ab}\alpha^{cd}\eta_{ca}v_d\\
                    &=-12\gamma\alpha + (1)^{|\alpha|}6\gamma^2[\alpha,\gamma]_V.
                \end{align*}
            Now, since $[\alpha,\gamma^3]_V=3\gamma^2[\alpha,\gamma]_V-(1)^{|\alpha|}12\gamma\alpha$, one has that 
                \begin{equation*}
                    \gamma^3\gamma^a\gamma^b [v_a,[v_b,\alpha]]=(1)^{|\alpha|}2[\alpha,\gamma^3]_V + 12\gamma\alpha,
                \end{equation*}
            hence 
                \begin{equation*}
                    \bar{\chi}\gamma^3[\alpha,\psi]=3\bar{\chi}\gamma\psi + (-1)^{|\alpha|}\frac{1}{2}\bar{\chi}[\alpha,\gamma^3]_v\psi.
                \end{equation*}
            \item \eqref{flip:0}. We use the fact that $C^t=-C$, hence
                \begin{align*}
                    \bar{\chi}\psi&=\chi^\alpha C_{\alpha\beta} \psi^\beta=(-1)^{|\chi||\psi|}\psi^\beta C_{\alpha \beta} \chi^\alpha = -(-1)^{|\chi||\psi|}\psi^\beta C_{\beta\alpha } \chi^\alpha = -(-1)^{|\chi||\psi|} \bar{\psi}\chi;
                \end{align*}
            
            \item \eqref{flip:1}. We denote by $(\Gamma^a)^\cdot_{\alpha \beta}:=C_{\alpha \delta}(\Gamma^a)^\delta_\beta$ and by $(\Gamma^a)_{\alpha \beta}=\delta_{\alpha \epsilon} (\Gamma^a)^\epsilon_\beta$. Then, using  $C\Gamma^a=-(\Gamma^a)^t C$, we have
                \begin{align*}
                    (\Gamma^a)^\cdot_{\alpha \delta}&=C_{\alpha \beta}(\Gamma^a)_\delta^\beta=\delta_{\alpha \epsilon} C^\epsilon_\beta (\Gamma^a)^\beta_\delta = -\delta_{\alpha \epsilon} (\Gamma^{a,t})^\epsilon_\beta C^\beta_\delta= -(\Gamma^{a,t})_{\alpha \beta} C^\beta_\delta \\
                    &=-(\Gamma^a)_{\beta\alpha}C^\beta_\delta=-(\Gamma^a)^\epsilon_\alpha \delta_{\epsilon \beta} C^\beta_\delta=-C_{\epsilon \delta}(\Gamma^a)^\epsilon_\alpha = C_{ \delta\epsilon}(\Gamma^a)^\epsilon_\alpha\\
                    &= (\Gamma^a)^\cdot_{\delta \alpha},
                \end{align*}
            hence finding 
                \begin{equation}
                    C_{\alpha\beta}(\Gamma^a)^\beta_\delta=-C_{\beta\delta}(\Gamma^a)^\beta_\alpha=C_{\delta\beta}(\Gamma^a)^\beta_\alpha.
                \end{equation}
            Now we have    
                \begin{align*}
                    \bar{\chi}\Gamma \psi &= (-1)^{|\psi|} \bar{\chi}^\alpha C_{\alpha\beta}(\Gamma^a )^\beta_\delta \psi^\delta v_a = (-1)^{|\psi|} \bar{\chi}^\alpha (\Gamma^a)^\cdot_{\alpha \beta} \psi^\beta v_a \\
                    & = (-1)^{|\psi|+|\psi||\chi|}\psi^\beta (\Gamma^a)^\cdot_{\beta\alpha }\chi^\alpha v_a=(-1)^{|\chi|+|\psi|+|\psi||\chi|}\bar{\psi}\Gamma\chi;
                \end{align*}
            \item \eqref{flip:2}. Recall $\Gamma^{ab}:=\Gamma^{[a}\Gamma^{b]}=\frac{1}{2}[\Gamma^a,\Gamma^b].$ Now
                \begin{align*}
                    (\Gamma^a\Gamma^b)^\cdot_{\alpha\beta}&=(\Gamma^a)^\cdot_{\alpha\delta}(\Gamma^b)^\delta_\beta=(\Gamma^a)^\cdot_{\delta\alpha}(\Gamma^b)^\delta_\beta=C_{\delta\epsilon}(\Gamma^a)^\epsilon_\alpha(\Gamma^b)^\delta_\beta\\
                    &=-C_{\epsilon\delta}(\Gamma^a)^\epsilon_\alpha(\Gamma^b)^\delta_\beta=-(\Gamma^b)^\cdot_{\epsilon\beta}(\Gamma^a)^\epsilon_\alpha=-(\Gamma^b)^\cdot_{\beta\epsilon}(\Gamma^a)^\epsilon_\alpha\\
                    &=-(\Gamma^b\Gamma^a)^\cdot_{\beta\alpha},
                \end{align*}
            implying $(\Gamma^{ab})^\cdot_{\alpha\beta}=- (\Gamma^{ba})^\cdot_{\beta\alpha}=(\Gamma^{ab})^\cdot_{\beta\alpha}$, finding
                \begin{align*}
                    \bar{\chi}\Gamma^2 \psi=\bar{\chi}^\alpha(\Gamma^{ab})^\cdot_{\alpha\beta} \psi^\beta v_a v_b = (-1)^{|\psi||\chi|}\bar{\psi}^\beta(\Gamma^{ab})^\cdot_{\beta\alpha}\chi^\alpha v_av_b=(-1)^{|\psi||\chi|}\bar{\psi}\Gamma^2\chi;
                \end{align*}
            \item \eqref{flip:3}. Again $\Gamma^{abc}=\Gamma^{[a}\Gamma^b\Gamma^{c]}$, and
                \begin{align*}
                    (\Gamma^a\Gamma^b\Gamma^c)^\cdot_{\alpha\beta}&=(\Gamma^a\Gamma^b)^\cdot_{\alpha\delta}(\Gamma^c)^\delta_\beta=-(\Gamma^b\Gamma^a)^\cdot_{\delta\alpha}(\Gamma^c)^\delta_\beta=-C_{\delta\epsilon}(\Gamma^b\Gamma^a)^\epsilon_\alpha(\Gamma^c)^\delta_\beta\\
                    &=C_{\epsilon\delta}(\Gamma^b\Gamma^a)^\epsilon_\alpha(\Gamma^c)^\delta_\beta=(\Gamma^c)^\cdot_{\epsilon\beta}(\Gamma^b\Gamma^a)^\epsilon_\alpha=(\Gamma^c)^\cdot_{\beta\epsilon}(\Gamma^b\Gamma^a)^\epsilon_\alpha\\
                    &=(\Gamma^c\Gamma^b\Gamma^a)^\cdot_{\beta\alpha},
                \end{align*}                
            implying $(\Gamma^{abc})^\cdot_{\alpha\beta}=-(\Gamma^{abc})^\cdot_{\beta\alpha} $, which in turn gives
                \begin{align*}
                    \bar{\chi}\Gamma^3\psi&=(-1)^{|\psi|}\chi^\alpha(\Gamma^{abc})^\cdot_{\alpha\beta}\psi^\beta v_a v_b v_c = -(-1)^{|\psi| +|\psi||\chi| }\psi^\beta(\Gamma^{abc})^\cdot_{\beta\alpha}\chi^\alpha v_av_bv_c \\
                    &=  -(-1)^{|\psi| +|\chi| +|\psi||\chi| }\bar{\psi}\Gamma^3 \chi;
                \end{align*}
            \item \eqref{flip:N}. In order to prove the general formula, we first have to prove 
                \begin{equation*}
                    (\Gamma^{a_1\cdots a_r})^\cdot_{\alpha\beta}=-t_r(\Gamma^{a_1\cdots a_r})^\cdot_{\beta\alpha}.
                \end{equation*}
            In particular, we want to show that $t_r=(-1)^{\floor*{\frac{r+1}{2}}}$. We know this is true for $r=0,1,2,3$ as we showed explicitly the values of $t_r$ in these cases. Now we prove the inductive step. Consider
                \begin{align*}
                    (\Gamma^{a_1\cdots a_r} \Gamma^{a_{r+1}})^\cdot_{\alpha\beta}&=(\Gamma^{a_1\cdots a_r} )^\cdot_{\alpha\delta}(\Gamma^{a_{r+1}})^\delta_\beta=-(-1)^{\floor*{\frac{r+1}{2}}}(\Gamma^{a_1\cdots a_r})^\cdot_{\delta\alpha}(\Gamma^{a_{r+1}})^\delta_\beta\\
                    &=(-1)^{\floor*{\frac{r+1}{2}}}C_{\epsilon\delta}(\Gamma^{a_1\cdots a_r})^\epsilon_\alpha(\Gamma^{a_{r+1}})^\delta_\beta=(-1)^{\floor*{\frac{r+1}{2}}}(\Gamma^{a_{r+1}})^\cdot_{\beta\epsilon}(\Gamma^{a_1\cdots a_r})^\epsilon_\alpha\\
                    &=(-1)^{\floor*{\frac{r+1}{2}}}(\Gamma^{a_{r+1}}\Gamma^{a_1\cdots a_r})^\cdot_{\beta\alpha}, 
                \end{align*}
            implying 
                \begin{align*}
                    (\Gamma^{a_1\cdots a_{r+1}})^\cdot_{\alpha\beta}&=(-1)^{\floor*{\frac{r+1}{2}}}(\Gamma^{a_{r+1}a_1\cdots a_r})^\cdot_{\beta\alpha}=(-1)^{\floor*{\frac{r+1}{2}}+r}(\Gamma^{a_1\cdots a_{r+1}})^\cdot_{\beta\alpha}\\
                    &=-(-1)^{\floor*{\frac{r+1}{2}}+r+1}(\Gamma^{a_1\cdots a_{r+1}})^\cdot_{\beta\alpha}=-(-1)^{\floor*{\frac{r+2}{2}}}(\Gamma^{a_1\cdots a_{r+1}})^\cdot_{\beta\alpha},
                \end{align*}
            showing that $t_{r+1}=(-1)^{\floor*{\frac{r+2}{2}}}$ as expected.\footnote{Here we used the fact that $(-1)^{\floor*{\frac{n}{2}}+n}=(-1)^{\floor*{\frac{n+1}{2}}}  $, as one can easily check by separating the cases for $n=2k,2k+1$. } With this formula, we can now easily show
                \begin{align*}
                    \bar{\chi}\Gamma^N\psi&=(-1)^{N|\psi|} \chi^\alpha((\Gamma^{a_1\cdots a_N})^\cdot_{\alpha\beta}\psi^\beta v_{a_1}\cdots v_{a_N}\\
                    &=-t_N(-1)^{N|\psi|+|\psi||\chi|}\psi^\beta(\Gamma^{a_1\cdots a_N})^\cdot_{\beta\alpha}\chi^\alpha v_{a_1}\cdots v_{a_N}\\
                    &=-t_N(-1)^{N(|\psi|+|\chi|)+|\psi||\chi|}\bar{\psi}\Gamma^N\chi
                \end{align*}
            \item \eqref{Fierz:1} and \eqref{Fierz:2}. We consider four Majorana spinors $\lambda_i$ of arbitrary parity. First we see that
                \begin{align*}
                    \bar{\lambda}_1\gamma^3\lambda_2\bar{\lambda}_3\gamma\lambda_4&=-(-1)^{|\lambda_2|+|\lambda_3|}\bar{\lambda}_1\gamma^{bcd}\lambda_2\bar{\lambda}_3\gamma^a\lambda_4 v_a v_b v_c v_d\\
                    &=-4!(-1)^{|\lambda_2|+|\lambda_3|}\bar{\lambda}_1\gamma^{bcd}\lambda_2\bar{\lambda}_3\gamma^a\lambda_4 \epsilon_{abcd}v_0v_1v_2v_3\\
                    &\overset{\eqref{id: gamma5gammac}}{=}-4! i (-1)^{|\lambda_2|+|\lambda_3|}\bar{\lambda}_1\gamma^5\gamma_a\lambda_2\bar{\lambda}_3\gamma^a\lambda_4 v_0v_1v_2v_3\\
                    &=4! i (-1)^{|\lambda_2|+|\lambda_3|}\bar{\lambda}_1\gamma_a\gamma^5\lambda_2\bar{\lambda}_3\gamma^a\lambda_4 v_0v_1v_2v_3,
                \end{align*}
            having used $\{\gamma^5,\gamma^a\}=0.$. Redefining $\lambda_2':=\gamma^5\lambda_2$ and $\bar{\lambda}_1':=\bar{\lambda}_1\gamma^5$ and setting $v^4=v_0v_1v_2v_3$, we obtain
                \begin{align}
                    \label{id:rearr_1}\bar{\lambda}_1\gamma^3\lambda_2\bar{\lambda}_3\gamma\lambda_4&=-4! i (-1)^{|\lambda_2|+|\lambda_3|}\bar{\lambda}'_1\gamma_a\lambda_2\bar{\lambda}_3\gamma^a\lambda_4 v^4\\
                    &=4! i (-1)^{|\lambda_2|+|\lambda_3|}\bar{\lambda}_1\gamma_a\lambda'_2\bar{\lambda}_3\gamma^a\lambda_4 v^4
                    \label{id:rearr_2}
                \end{align}
            We now apply \eqref{Fierz:0} to the  expressions containing $\gamma^a\gamma_a$. Explicitly
                \begin{align*}
                    3\lambda_1'^\alpha\lambda_2^\beta\lambda_3^\rho\lambda_4^\delta (\gamma^a)^\cdot_{\alpha (\beta}(\gamma_a)^\cdot_{\rho \delta)} &= \lambda_1'^\alpha\lambda_2^\beta\lambda_3^\rho\lambda_4^\delta( (\gamma^a)^\cdot_{\alpha \beta}(\gamma_a)^\cdot_{\rho \delta} + (\gamma^a)^\cdot_{\alpha \rho}(\gamma_a)^\cdot_{\beta \delta} +(\gamma^a)^\cdot_{\alpha \delta}(\gamma_a)^\cdot_{\beta\rho } )   \\
                    &=\bar{\lambda}'_1\gamma^a\lambda_2\bar{\lambda}_3\gamma_a\lambda_4 +(-1)^{|\lambda_2||\lambda_3|}\bar{\lambda}'_1\gamma^a\lambda_3\bar{\lambda}_2\gamma_a\lambda_4\\
                    &\quad + (-1)^{|\lambda_4|(|\lambda_2|+|\lambda_3|)}\bar{\lambda}'_1\gamma^a\lambda_4\bar{\lambda}_2\gamma_a\lambda_3\\
                    &=0, 
                \end{align*}
            substituing in \eqref{id:rearr_1} gives
                \begin{align*}
                    \bar{\lambda}_1\gamma^3\lambda_2\bar{\lambda}_3\gamma\lambda_4&=-i\cdot4!(-1)^{|\lambda_2|+|\lambda_3|}[ (-1)^{|\lambda_2|+|\lambda_3|} \bar{\lambda}_1\gamma^5\gamma_a\lambda_3\bar{\lambda}_2\gamma^a\lambda_4 \\
                    &\quad +  (-1)^{|\lambda_4|(|\lambda_2|+|\lambda_3|)}\bar{\lambda}_1\gamma^5\gamma_a\lambda_4\bar{\lambda}_2\gamma^a\lambda_3]v^4\\
                    &\overset{\eqref{id: gamma5gammac}}{=}-(-1)^{|\lambda_2||\lambda_3|}\bar{\lambda}_1\gamma^3\lambda_3\bar{\lambda}_2\gamma\lambda_4 - (-1)^{|\lambda_3|+|\lambda_4|(|\lambda_2|+|\lambda_3|+1)}\bar{\lambda}_1\gamma^3\lambda_4\bar{\lambda}_2\gamma\lambda_3.
                \end{align*}
            \eqref{Fierz:1} is recovered in the same way applying \eqref{Fierz:0} to \eqref{id:rearr_2}.   
        \end{itemize}


\newrefcontext[sorting=nty]
\sloppy
\printbibliography   

\end{document}